\documentclass[a4paper,twoside,12pt]{article}

\usepackage{graphicx}
\usepackage[body={17.5cm, 22cm},right=2cm]{geometry}
\usepackage{amssymb}

\newcommand\ee{\end{equation}}
\newcommand\be{\begin{equation}}
\newcommand\eea{\end{eqnarray}}
\newcommand\bea{\begin{eqnarray}}

\newcommand\de{\partial}
\newcommand{\sfrac}[2]{{\textstyle\frac{#1}{#2}}}
\newcommand\Lc{\Lambda^3}
\newcommand\pit{\hat\pi}

\begin{document}

\begin{flushright}
IFT-UAM/CSIC-04-13\\
CERN-PH-TH/2004-063\\
\end{flushright}

\vspace{10pt}

\begin{center}
{\LARGE \bf Classical and Quantum Consistency \\ [0.3cm]
of the DGP Model}

\vspace{20pt}

{Alberto Nicolis$\, ^a$ and Riccardo Rattazzi$\, ^b$} 

\vspace{15pt}

$^a$\textit{Instituto de F\'\i sica Te\'orica, C--XVI, UAM, 28049 Madrid, Spain}

\vspace{5pt}

$^b$\textit{Physics Department, Theory Division, CERN, CH--1211 Geneva 23, Switzerland}
\end{center}

\vspace{5pt}

\abstract{ We study the Dvali-Gabadadze-Porrati model by the method of the boundary effective action.
The truncation of this action to the bending mode $\pi$ consistently describes physics in
a wide range of regimes both at the classical and at the quantum level. The Vainshtein effect,
which restores agreement with precise tests of general relativity, 
follows straightforwardly.
We give a simple and general proof
of  stability,  i.e.~absence of ghosts  in the fluctuations, valid for
most of the relevant cases, like for instance the spherical source in asymptotically flat space. 
However we confirm that around certain interesting self-accelerating cosmological solutions there is a ghost.
We consider the issue of quantum corrections. Around flat space $\pi$ becomes strongly coupled 
below a macroscopic length of 1000 km, thus impairing the predictivity of the model. Indeed 
the tower of higher dimensional operators which is expected by a generic UV completion of the model
limits predictivity at even larger length scales. We outline a non-generic but consistent choice
of counterterms for which this disaster does not happen and for which the model remains calculable 
and successful in all the  astrophysical situations of interest. By this choice, the
extrinsic  curvature $K_{\mu\nu}$ acts roughly like a dilaton field controlling the strength
of the interaction and the cut-off scale at each space-time point.
At the surface of Earth the cutoff is $\sim 1$ cm
but it is unlikely that the associated  quantum effects be observable in table
 top experiments.}


\section{Introduction}

In recent years several attempts have been made to formulate a theoretically consistent extension of General 
Relativity that modifies gravity at cosmological distances while remaining compatible with observations at shorter distances.
Well known examples in the literature are models of massive gravity \cite{PF,van}, the GRS model \cite{GRS}, the DGP model 
\cite{DGP}, and the newborn model of `ghost condensation' \cite{ACLM}.
These attempts are motivated both by the appealing theoretical challenge they represent, and, perhaps more 
interestingly, by the experimental indication that something strange is indeed happening at very large scales:
the expansion of the Universe is accelerating. 
This fact can be explained, as it is usually done, by invoking a small cosmological constant, or some sort of `dark
energy'. Nevertheless, it is also worth considering the possibility that at cosmological 
distances gravity itself is different from what we experience at much smaller scales, in the solar system.
This perspective could perhaps offer a new direction for addressing the cosmological constant problem (see for instance ref.~\cite{Dvali:2002pe}).

In the present paper we concentrate on the DGP model. 
In this model there exists a critical length scale $L_{\rm DGP}$ below which gravity looks
four dimensional, while for larger distances it weakens and becomes five dimensional.
To avoid conflict with observations, the parameters of the model can be adjusted to make $L_{\rm DGP}$ of the order 
of the present Hubble horizon $H_0^{-1}$.
One cosmological solution of DGP is an ordinary Friedmann-Robertson-Walker Universe that, even in the absence 
of a cosmological constant, gets accelerated at late times by the weakening of gravity
\cite{deffayet,DDG}. Moreover, even though the length scale of gravity modification is cosmological,
one still expects small but non-negligible
effects on solar system astrophysical measurements \cite{LS,DGZ}.
This makes the model extremely interesting from the phenomenological point of view.

In spite of these appealing features, the model is plagued by some consistency problems, as pointed out in 
ref.~\cite{LPR}.
There it was shown that the theory has strong interactions at a tiny energy scale
\be 
\Lambda \sim (M_P / L_{\rm DGP}^2)^{1/3} \; ,
\label{thousand}
\ee
which corresponds to wavelengths of about 1000 km (a qualitatively similar, but quantitatively different, result was found
in ref.~\cite{rub}). 
This means that below 1000 km the model loses predictivity, and one cannot perform sensible computations 
without knowing the UV completion of the theory.
In the presence of curvature the situation seems even worse. Indeed in  a non-renormalizable theory like the one at hand, 
quantum corrections necessarily imply the presence of an infinite tower of higher dimensional operators
suppressed by inverse powers of the  interaction scale $\Lambda$.  A  `generic'
choice of this tower of operators  disrupts the classical computations of the original DGP model \cite{LPR},
making it practically uncalculable, whenever the Riemann curvature is bigger than ${\cal O}(1/L_{\rm DGP}^2)$.
In practice the unknown UV completion of the model dominates physics above the tiny curvature $1/L_{\rm DGP}^2$.
In the case of a spherical source in asymptotically flat space 
this happens below a distance from the source  given by  the so called
Vainshtein length \cite{v,ddgv}
\be
R_V \sim (R_S \, L_{\rm DGP}^2)^{1/3} \; ,
\ee
where $R_S$ is the Schwarzschild radius of the source.
For instance, for the Sun $R_V$ is about $10^{20}$ cm! Apparently one cannot compute
the gravitational potential of the Sun at distances shorter than $10^{20}$ cm.
A completely equivalent problem has been pointed out in the case of massive gravity \cite{AGS}.
It should be stressed that, in the purely classical DGP model, $R_V$ is precisely the  length below which
classical non-linearities in the field equations eliminate the phenomenologically unwanted effects of the
extra polarizations of the graviton \cite{v,ddgv}. 
The latter are usually referred to as the van Dam-Veltman-Zakharov discontinuity \cite{vdvz}. These non-linearities
are therefore a virtue of the classical theory, but upon quantization they suggest the presence of many more uncontrollable
interactions.

Finally, the model possesses negative energy classical solutions, which signal the presence of instabilities
already at the classical level. These negative energy solutions have however a typical curvature larger than $1/L_{\rm DGP}^2$,
so that their true presence, or absence, is fully dependent on the UV completion.

In this paper we try to understand to what extent all these results really threaten the consistency and the 
predictivity of the model. In spite of the above difficulties we will be able to draw fairly optimistic conclusions.
In practice we will show that one can consistently assume a UV completion ($\equiv$ tower of higher dimensional operators)
where the theory can be extrapolated down to distances significantly shorter than $1/\Lambda$. Though our choice is 
consistent, in the sense that it is stable under RG evolution,
it  may at first sight look unnatural from the effective field theory view point. 
We will try to argue that perhaps it  isn't.

In the DGP model our world is the 4D boundary of an infinite 5D spacetime. 
One can integrate out the bulk degrees of freedom and find an ``effective'' action for the 4D fields. 
This was explicitly done in ref.~\cite{LPR}, where it was found that from the 4D point of view, beside the 
ordinary graviton, there is an extra scalar degree of freedom $\pi$ that plays a crucial role.
This is essentially a brane bending mode contributing to the extrinsic curvature of the boundary like 
$K_{\mu\nu}\propto \partial_\mu\partial_\nu \pi$. 
It is this scalar that interacts strongly at momenta of order $\Lambda$.
Indeed there exists a limit in which the strong interaction scale is kept fixed but all other degrees of freedom 
(namely the graviton and a vector $N_\mu$) decouple.
This means that in order to grasp the interesting physics related to the strong interaction we can
restrict to the $\pi$ sector, which represents a consistent truncation of the theory. We will see
that all the interesting features (good or bad) of the DGP model can be traced to the dynamics of this scalar $\pi$.
For our purposes we can therefore forget about the 5D geometric setup of the model and study a scalar
theory with a specific cubic interaction, namely $\frac 1 \Lc (\de \pi)^2 \, \Box \pi$.
This largely simplifies the analysis.

We start in sect.~\ref{general} by studying classical field configurations in this theory.
For any given distribution of matter sources, the field equation for $\pi$
turns out to be a quadratic {\em algebraic} equation for the tensor of its second 
derivatives $\de_\mu \de_\nu \pi$.
Like ordinary quadratic equations, it possesses doublets of solutions: for any solution $\pi^+ (x)$, there exists 
a corresponding `conjugate' solution $\pi^-(x) \neq \pi^+ (x)$.
By studying the stability of a generic solution against small fluctuations, one discovers
that this conjugation exactly reverses the sign of the kinetic action of the fluctuations.
This means that if $\pi^+ (x)$ is a stable solution, its conjugate $\pi^-(x)$ must be unstable.
It is therefore not surprising that classical instabilities were found in ref.~\cite{LPR}.

Does the existence of these unstable solutions make the model inconsistent at the classical level? 
We argue that this is not the case: the classical stability of the $\pi$ system is guaranteed for 
a well defined class of energy-momentum sources and boundary conditions at spatial infinity, as we prove explicitly in 
sect.~\ref{stability}. In particular, we show that for any configuration of sources with
positive energy density decaying at spatial infinity, and small pressure (smaller than $\rho/3$),
the solution which is trivial at spatial infinity is stable everywhere\footnote{If the sources have 
relativistic velocities an additional hypothesis is required, see sect.~\ref{stability}.}.
Of course the conjugate configuration, which instead will diverge at infinity, will be unstable.
The point is that the sets of stable and unstable solutions are disconnected in field space, they cannot
`communicate'. It suffices that a solution is locally stable at one space-time point to ensure
its global stability in the whole Universe, because there exists no continuous path of solutions 
that connects the stable and the unstable regions in field space.
Unstable solutions simply correspond to the `wrong' choice of boundary conditions.

Unfortunately nothing prevents some interesting solutions from having the wrong boundary behaviour: it
is the case of the de Sitter solution, which describes the self-accelerated Universe we mentioned above.
In ref.~\cite{LPR} it was pointed out that such solution is plagued by ghost-like instabilities. 
In sect.~\ref{deSitter} we check this result in our formalism, indeed finding that the self-interacting dynamics
of $\pi$ is responsible for the existence of the de Sitter solution. The latter turns out to be
nothing but the conjugate of the trivial configuration $\pi (x) = 0$, and as such must be unstable.

In sect.~\ref{spherical} we study the solution for $\pi$ in presence of a spherical heavy source.
We reproduce in the $\pi$ language the results of refs.~\cite{g,p3,LS,Tanaka}, 
namely that the correction  induced by the DGP model scales like $(r/R_V)^{3/2}$ relative to Newton's law. As stressed in ref.~\cite{LS},
the sign of this correction depends on
the cosmological phase, i.e.~on the behaviour of the Universe at infinity. 
In accordance with the general proof given in sect.~\ref{stability}, we explicitly check that
fluctuations around this solution are stable.
Also in this case the interesting features of DGP are encoded in the non trivial dynamics of the field $\pi$,
in a simple and transparent way. 

The results of sects.~\ref{stability}, \ref{deSitter}, \ref{spherical} imply that for any astrophysical source and
for a wide and relevant class of cosmological solutions no classical instability develops in the $\pi$ field.
Finally, in sect.~\ref{quantum}, we discuss the problem of quantum corrections. We have already
mentioned that a generic choice of higher dimension operators leads to uncalculability when
 the Riemann curvature is bigger than ${\cal O}(1/L_{\rm DGP}^2)\sim H_0^2$. We look at this problem in more detail.
We characterize higher order corrections, distinguishing between a classical and a quantum expansion parameter.
We notice  that the large effects of the counterterms which were 
pointed out is ref.~\cite{LPR} are more related to the classical 
expansion that to the quantum one. 
 We clarify this, by  discussing first the structure of the 1-loop effective action.
We show  that  the terms which are truly needed for unitarity and consistency, 
and which are associated to logarithmic divergences, do not grow in the regions of large curvature. 
In a sense  this follows simply from the fact that the classical theory is well defined, i.e.~there are no ghosts.
Inspired by this simple example we then proceed to give a general and consistent characterization
of the class of counterterm Lagrangians  which do not  destabilize the naive classical results. Our claim
is based on a remarkable property of the tree level action for  $\pi$. The large curvature backgrounds
around which one seemingly looses calculability 
correspond to a classical background  $|\partial_\mu\partial_\nu \pi_{cl}/\Lambda^3|\gg 1$. Around such a background
 the quantum fluctuation $\varphi=\pi-\pi_{cl}$ receives a
 large positive contribution to its kinetic term
\be
{\cal L}_{\rm kin}=Z_{\mu\nu}(\partial^\mu \varphi) (\partial^\nu\varphi) \; ,
\ee
where $Z_{\mu\nu}$ is a non singular matrix proportional to $\partial_\mu\partial_\nu \pi_{cl}/\Lambda^3$ itself. The
interaction Lagrangian remains however the same: no large factors there. The large kinetic term then suppresses interactions
(as seen for instance by going to canonical fields).
At each  spacetime point $x$ one can define (suppressing spacetime indices) a naive local scale 
$\tilde \Lambda(x)=\Lambda {\sqrt {Z(x)}}\gg \Lambda$,
which is a function of the background field $\pi_{cl}$ itself.
Our basic result is then the following. If the counterterm action  depends on $\pi$ and $\Lambda$ {\it only} 
via the running scale
$\tilde \Lambda$ and its derivatives, then the classical DGP action can be used down to the distance $1/\tilde \Lambda(x)$
at each point $x$. This implies, in particular, that the Vainshtein solutions within the solar system are always fine.
On the other hand at the Earth surface one finds $\tilde \Lambda\sim 1 \: {\rm cm}^{-1}$. In principle this could be worrisome,
or interesting. In practice it is probably neither. This is because at the surface of Earth 
$\pi$ is well decoupled from matter.

The counterterm structure we select must correspond to some specific UV completion for the physics
at the 4D boundary. At the scale $\tilde \Lambda$ we expect new states to become relevant. This scale
is determined by the background expectation value of the massless field $\pi$, or more precisely $\partial_\mu\partial_\nu \pi
\propto K_{\mu\nu}$. Therefore the extrinsic curvature acts like a sort of dilaton controlling the scale of the new physics.
A direct consequence of this property is  that our ansatz can be motivated by an approximate
scale invariance of the UV completion. 

We conclude with an outlook on possible developments of our results.


\section{Boundary effective action}\label{effective}

The DGP model \cite{DGP} describes gravity in a five-dimensional spacetime ${\cal{M}}$, with boundary $\de {\cal{M}}$.
The action is postulated to be purely Einstein-Hilbert in the bulk, plus a four dimensional 
Einstein-Hilbert term localized at the boundary, with different Planck scales,
\be
S_{\rm DGP} = 2 M^3_5 \int_{\cal{M}} \! d^5 x \sqrt{-G} \, R(G) 
+ 2 M^2_4 \int_{\de \cal{M}} \! d^4 x \sqrt{-g} \, R (g) - 4 M^3_5 \int_{\de \cal{M}} \! d^4 x \sqrt{-g} \, K (g) \; ,
\ee
where $G_{MN}$ is the 5D metric, $g_{\mu\nu}$ is the 4D induced metric on the boundary,
and we have added the Gibbons-Hawking term at the boundary in order to obtain the 5D Einstein equations
upon variation of the bulk action \cite{GH}.
Matter fields and a possible cosmological constant are supposed to be localized on the boundary.
A special role is played by the length scale $L_{\rm DGP} = 1/m \equiv M^2_4 / M^3_5 $: below
$L_{\rm DGP}$ gravity looks four-dimensional, while at larger length scales it enters in the five-dimensional
regime. 
In order for the model to be viable, $L_{\rm DGP}$ must be huge, at least of the order of the present Hubble horizon.

In the following sections we will refer to the 4D boundary effective action obtained in ref.~\cite{LPR}
by integrating out the bulk degrees of freedom. 
We summarize here only the main results.
By using spacetime coordinates $\{x^\mu,y\}$, with the boundary sitting at $y=0$, the bulk piece of the action
can be rewritten in terms of ADM-like variables as
\be \label{bulk}
S_{\rm bulk} = 2 M^3_5 \int \! d^4 x \int_0 ^\infty \! dy 
\sqrt{-g} \, N \left[ R(g) - K^{\mu\nu} K_{\mu\nu} + K^2
\right] \; ,
\ee
where $N=1/\sqrt{G^{yy}}$ is the lapse, $N_\mu = G_{y\mu}$ is the shift, $g_{\mu\nu} = G_{\mu\nu}$ is the metric
on surfaces of constant $y$, and the extrinsic curvature tensor is
\be \label{extrinsic}
K_{\mu\nu} = \frac 1 {2N} \left( \de_y g_{\mu\nu} -\nabla_\mu N_\nu -\nabla_\nu N_\mu \right) \; .
\ee
One then expands the 5D metric and all the other geometric quantities around
a flat background, $G_{MN} = \eta_{MN} + h_{MN}$, and integrate out the bulk in order to obtain an
effective action for the 4D dimensional fields living on the boundary.
The final result, at the quadratic level, is
\be \label{kinetic}
S_{\rm bdy} \simeq M^2_4 \int \! d^4 x \, \left[ \sfrac 1 2 \, h'^{\mu\nu} \, \Box  h'_{\mu\nu} - \sfrac 1 4 \, 
h' \, \Box  h'  - m \,  N'^\mu \Delta  N'_\mu + 3 m^2 \, \pi \, \Box \pi
\right]  \; ,
\ee
where $\Delta$ is a non-local differential operator, $\Delta = \sqrt{-\Box} = \sqrt{-\eta^{\mu\nu} \, \de_\mu \de_\nu}$,
and the kinetic terms have been diagonalized by defining
\be \label{diagonal}
h_{yy} = -2 \Delta \pi \; , \qquad N'_\mu = N_\mu - \de_\mu \pi \; , 
\qquad h'_{\mu\nu} = h_{\mu\nu} - m \, \pi \, \eta_{\mu\nu} \; . 
\ee

By taking into account bulk interactions with higher powers of $h_{MN}$, one finds that the leading 
boundary interaction term is cubic in $\pi$, and involves four derivatives,
\be \label{interaction}
\Delta S^{(3)}_{\rm bdy} = - M^3_5 \int \! d^4 x \, (\de \pi)^2 \, \Box \pi \; .
\ee
The comparison of this interaction with the kinetic term of the $\pi$ field in eq.~(\ref{kinetic}) immediately 
shows that the $\pi$
sector of the theory becomes strongly interacting at a very small energy scale 
$\Lambda = (m^2 M_4)^{1/3} = M^2_5 / M_4$, which corresponds to a length scale of about 1000 km.

The boundary effective action we will use throughout the paper is the sum $S_{\rm bdy}+ \Delta S^{(3)}_{\rm bdy}$.
Notice that $\Delta S^{(3)}_{\rm bdy}$ is the leading boundary interaction term in a quantum sense: it 
gives the largest amplitude
in low energy scattering processes, and the scale $\Lambda$ associated to it is the lowest of all strong
interaction scales associated to further interaction terms. 
These terms are schematically of the form \cite{LPR}
\be  \label{general_int}
\Delta {\cal{L}}_{\rm bdy} \sim M_5^3 \, \de \, (N'_\mu)^p \, (\de \pi)^{q} \, (h'_{\mu\nu})^s 
\sim m M^2_4 \, \de \, \left( \frac{\hat N'_\mu}{m^{1/2} M_4} \right)^p \, 
\left( \frac{\de \hat \pi}{m M_4}\right)^{q} \, (h'_{\mu\nu})^s   \; ,
\ee
where $\hat N'_\mu$ and $\hat \pi$ are canonically normalized, and $p+q+s \ge 3$. 
Eq.~(\ref{interaction}) corresponds to the term with $p=s=0$ and $q=3$.

However, in this paper we study classical field configurations, and from a classical point of view
it is not immediately manifest why eq.~(\ref{interaction}) should be the most important interaction term, and
why we should be allowed to neglect all the others. 
In fact, the strong interaction scale $\Lambda$ has no meaning at the classical level (see discussion in sect.~\ref{quantum}).
Nevertheless, we will check explicitly in the two specific cases we will deal with 
(the de Sitter solution and the spherically symmetric solution) that indeed all other 
interaction terms are subdominant as long as the flat-space approximation is valid, 
i.e.~as long as $|h'_{\mu\nu}| \ll 1$. 
This fact can be understood in a general although formal way. 
First, if $|h'_{\mu\nu}| \ll 1$, in eq.~(\ref{general_int}) we can stick to terms with $s=0$. 
Then, suppose we have a compact source for our fields: 
the region in which space is nearly flat is the region well outside the Schwarzschild radius 
$R_S$. 
We can mimic to be in this region by formally sending $R_S$ to zero, i.e.~by decoupling 
four-dimensional gravity from the source. 
In this limit the solution for the metric is trivial, $h'_{\mu\nu}=0$.
However, we do not want to end up in a completely trivial configuration, so we want to preserve the 
self-coupling of the $\pi$ field, as well as its interaction with the source: 
in terms of the canonically normalized $\hat \pi$ field,
the cubic self coupling is unchanged if we keep $\Lambda = M^2_5 / M_4$ fixed; 
in order to take into account the interaction with matter sources, we notice that
the interaction Lagrangian between the 4D metric perturbation $h_{\mu\nu}$ and the matter 
stress energy tensor $T_{\mu\nu}$ is, by definition, $\frac 1 2 h_{\mu\nu} T^{\mu\nu}$.
From the definition of $\pi$, eq.~(\ref{diagonal}), we see that $\hat \pi$ interacts with 
matter via a term $\frac 1{2 M_4} \hat \pi T$.
Therefore, if we take the formal limit 
\be \label{limit}
M_4, M_5, T_{\mu\nu} \to \infty \; , \qquad \frac{M^2_5}{M_4}= {\rm const} \, , 
\; \frac{T_{\mu\nu}}{M_4} = {\rm const} \; ,
\ee
we decouple 4D gravity while keeping the full Lagrangian for $\hat \pi$ fixed. By `full' we mean 
the sum of kinetic, cubic and source terms. We stress that when applying this limit to a
spherical source of mass $M$, the Vainshtein radius $R_V^3=R_S L_{\rm DGP}^2=MM_5^6/M_4^3$
remains fixed, showing that we are focusing on the genuine non-linearities of the DGP model. The point now is that in the above limit all
coefficients of further interactions, eq.~(\ref{general_int}), vanish,
\be
m M^2_4 \, \de \, \left( \frac{\hat N'_\mu}{m^{1/2} M_4} \right)^p \, 
\left( \frac{\de \hat \pi}{m M_4}\right)^{q} = 
\left( \frac{M_4}{M^2_5}\right)^q \frac 1{M_5^{3p/2+q-3}}\, 
\de \, (\hat N'_\mu)^p (\de \hat \pi)^q \to 0 \; .
\ee
So far we neglected the dynamics of $\hat N'_\mu$, but the above result tells us that we were 
justified in doing so: $\hat N'_\mu$ does not interact directly with matter, 
and in the limit we are considering all its interactions vanish, thus decoupling it from all 
other degrees of freedom. 
In conclusion, we see that in the limit in which the metric perturbation can be neglected, 
$|h'_{\mu\nu}| \ll 1$, eq.~(\ref{interaction}) gives the dominant interaction term also for
classical field configurations. Conversely our approximation breaks down when $|h'_{\mu\nu}|\sim 1$, like
when approaching a black-hole horizon. In fact when this happens we also have $(\partial \hat \pi/mM_4)\equiv \partial \pi\sim 1$ (see sect.
\ref{spherical}). Since $\pi$ roughly measures the bending of the brane, 
this simply tells us that our approximation breaks down when the brane can no longer be treated as approximately  flat in 5D. In the final section we will briefly explain
what we expect to happen in this regime.


\section{General properties of classical solutions}\label{general}

We now move to study classical solutions in presence of matter sources on the boundary.
We restrict to the $\pi$ sector, with cubic self-interactions and coupled to matter.
As we showed in the previous section, this is a consistent truncation of the theory, 
in the sense that there exist a limit, eq.~(\ref{limit}), in which all other degrees of freedom decouple,
and all further interactions vanish.


From the results reported in the previous section we have that the full action for
the $\pi$ field in flat space is
\be \label{action}
S = \int \! d^4 x \left[ - 3 (\de  \pit)^2 - \frac{1}{\Lc} (\de \pit)^2 \, {\Box} \pit + \frac{1}{2 M_4} \, \pit T
  \right]\;,
\ee 
where $T =T^\mu {}_\mu$, and we canonically normalized $\pi$.

The first variation of $S$ with respect to $\hat \pi$ gives the field equation
\be \label{eq_pi}
6 \, \Box \pit - \frac{1}{\Lc} \Box (\de \pit)^2 + \frac{2}{\Lc} \de_\mu (\de^\mu \pit \, \Box \pit) +  
\frac{T}{2 M_4} = 0 \; ,
\ee
which can be rewritten as
\be \label{GC_pi}
3 \, \Box \pit - \frac{1}{\Lc}(\de_\mu \de_\nu \pit)^2 + \frac{1}{\Lc} (\Box \pit)^2 = -\frac{T}{4 M_4} \; .
\ee
Notice that the terms involving third derivatives of $\pit$ have canceled out, thus leaving us with
an {\em algebraic} equation for its second derivatives.
In terms of the tensor field $\tilde K_{\mu\nu} (x) = - \frac { 1} \Lc \de_\mu \de_\nu \pit$ the above equation reads
\be \label{GC}
\tilde K ^2 - (\tilde K_{\mu\nu})^2 - 3 \tilde K =-\frac{T}{4 \Lc M_4} \;,
\ee 
where $\tilde K = \tilde K^\mu {}_\mu$.
This result is easy to understand from a geometric point of view, by referring to the 5D setup of the model.
The field equation that derives from varying the bulk action eq.~(\ref{bulk}) with respect to the lapse $N$ is
one of the Gauss-Codazzi equations,
\be
R(g) + K^{\mu\nu}K_{\mu\nu} - K^2 = 0 \; ,
\ee
which is an algebraic relation between intrinsic and extrinsic curvature of the boundary.
The intrinsic curvature $R(g)$ can be eliminated by means of the Einstein equations on the boundary,
\be
4 M^2_4 \, ( R_{\mu\nu}- \sfrac 1 2  \, g_{\mu\nu} \, R) - 4 M^3_5 (K_{\mu\nu}- g_{\mu\nu} \, K) = T_{\mu\nu} \; ,
\label{boundaryem}
\ee
thus obtaining an algebraic relation between the extrinsic curvature $K_{\mu\nu}$ and the matter stress-energy tensor.
The extrinsic curvature is easily related to $\pi$ by expanding eq.~(\ref{extrinsic}),
\be
K_{\mu\nu} = m \left[ - \frac {1} \Lc \de_\mu \de_\nu \pit \left[1
  + {\cal O}\left(\sfrac1{M_5} \cdot \sfrac{M_4}{M_5^2}\de \pit \right)\right]
  + {\cal O}\left(\sfrac 1 {M_5} \cdot \sfrac{M_4}{M_5^2} \de \hat h'_{\mu\nu} \right) 
  + {\cal O}\left(\sfrac1{M_5^{3/2}} \cdot \de \hat N'_\mu \right)\right]
\; .
\ee
In the limit we are considering $\pi$ gives the dominant contribution, 
$K_{\mu\nu} \simeq - \frac m \Lc \de_\mu \de_\nu \pit = m \tilde K_{\mu\nu}$, and this leads precisely
to our eq.~(\ref{GC}).
This remark also tells us that eq.~(\ref{GC}) is more fundamental than its counterpart written in terms of $\pit$,
eq.~(\ref{GC_pi}). 
The latter is valid only in the limit of nearly flat space, while the former is a combination of a geometric
identity and of the exact Einstein equations on the boundary, and as such is exact also in presence of large curvatures,
as long as the classical theory makes sense.
Of course in such a case the tensor $\tilde K_{\mu\nu}$ is the extrinsic curvature of the boundary
in units of $m$, rather than the second derivative of $\pit$.

Once a solution of eq.~(\ref{GC}) for a given distribution of sources has been found,
one may be interested in studying its stability.
In order to do that, it is necessary to perturb $\pit$ and 
expand the action up to quadratic order in the perturbation $\varphi$,
\be
S_\varphi  =  \int \! d^4 x \left[ - 3  (\de \varphi)^2-\frac{1}{\Lc} \left( \Box\pit\,
(\de \varphi)^2 
+2 \, \de_\mu \pit \, \de^\mu \varphi  \, \Box \varphi \right) \right] \, .
\ee
By using the identity $\de^\mu \varphi \Box \varphi = \de_\nu [\de^\nu \varphi \de^\mu \varphi
-\frac 1 2 \eta^{\mu \nu} (\de \varphi)^2]$, and integrating by parts, we find 
\bea
S_\varphi & = & \int \! d^4 x \left[ - 3 (\de \varphi)^2 + \frac{2}{\Lc} \left( \de_\mu \de_\nu \pit - \eta_{\mu\nu} 
\Box \pit \right) \, \de^\mu \varphi \de^\nu \varphi \right] = \label{2nd_variation} \\
& = & \int \! d^4 x \left[ - 3 (\de \varphi)^2 - 2 \left( \tilde K_{\mu \nu} - \eta_{\mu\nu} \tilde K \right)
  \, \de^\mu \varphi \de^\nu \varphi \right] \; .  \label{stable}
\eea
This result deserves some comments.
First, notice that the starting cubic interaction term involves four derivatives, but the resulting quadratic action for 
the fluctuation $\varphi$ does not contain terms with more than two derivatives acting on the $\varphi$'s. 
This remarkable feature makes the issue of stability more tractable, since it prevents the appearance
of ghost states at high momenta, when higher derivatives contributions to the quadratic action could become
important and destabilize the system. 
Second, the coefficients in $S_\varphi$ depend only on the tensor $\tilde K_{\mu\nu}$, in an algebraic way.
The fact that both   the field equations  and the condition for stability depend algebraically on  
$\tilde K_{\mu\nu}$ will play a crucial role in what follows.

A deeper insight into the structure of classical solutions  comes from the following remark.
For a given distribution of sources, eq.~(\ref{GC}) is a second-order algebraic equation for $\tilde K_{\mu\nu}$. 
Instead of having a single solution, it is reasonable to expect doublets of solutions 
$\tilde K^\pm_{\mu\nu}$, like in ordinary quadratic equations. 
Moreover, one expects that in general the stability of the system depends on the solution chosen inside
this doublet, since the quadratic Lagrangian for the fluctuation $\varphi$, eq.~(\ref{stable}), depends on 
$\tilde K_{\mu\nu}$.
Even better, given the simple dependence of the overall coefficient of $\de^\mu \varphi \de^\nu \varphi$ 
in eq.~(\ref{stable}) on $\tilde K_{\mu\nu}$,
\be \label{def_Z} 
Z_{\mu\nu} = - 3 \eta_{\mu\nu} - 2 ( \tilde K_{\mu\nu} - \eta_{\mu\nu} \tilde K ) \; ,
\ee
one can straightforwardly invert this relation and express $\tilde K_{\mu\nu}$ in terms of $Z_{\mu\nu}$,
\be 
\tilde K_{\mu\nu}= \sfrac12 \eta_{\mu\nu} - \sfrac12 ( Z_{\mu\nu} - \sfrac13 \eta_{\mu\nu} Z )\; ,
\ee
where $Z = Z^\mu {}_\mu$.
This is remarkable: now we can write our field equation, eq.~(\ref{GC}), directly
as an algebraic equation for the kinetic coefficient of the fluctuations,
\be \label{Z_GC}
\sfrac13 Z^2 - (Z_{\mu\nu})^2 = 6 - \frac {T} {2 \Lc M_4}\; .
\ee
In the above equation the linear term in $Z_{\mu\nu}$ has canceled out, thus leading to a purely quadratic 
equation. 
This explicitly shows that if $Z_{\mu\nu} ^{+}$ is a solution, so is 
its `conjugate' $Z_{\mu\nu} ^{-} \equiv -Z_{\mu\nu} ^{+}$.

By definition eq.~(\ref{stable}) now simply reads,
\be \label{Z_stable}
S_\varphi = \int \! d^4 x \, Z_{\mu \nu} (x)  \, \de^\mu \varphi \de^\nu \varphi \; ,
\ee
which shows that a solution is locally stable against small fluctuations if and only if the 
matrix $Z_{\mu\nu}$, once diagonalized, has the usual `healthy' signature $(+,-,-,-)$. 
(In the following section we will be more precise in defining the requirements that ensure the stability
of the system.)
Therefore, if a solution $Z_{\mu\nu} ^+$ is locally stable, its conjugate $Z_{\mu\nu} ^-$ is locally unstable. 
The kinetic terms of the small fluctuations around the two solutions are exactly opposite. 

Coming back to $\tilde K_{\mu\nu}$, we conclude that if $\tilde K_{\mu\nu} ^+$ is a solution for a given
matter distribution, then there always exists another solution 
$\tilde K_{\mu\nu} ^- = \eta_{\mu\nu} - \tilde K_{\mu\nu} ^+$ for the same matter distribution,
and the stability properties of the two solutions are opposite. 
Of course, since by definition $\tilde K_{\mu\nu} = -\frac 1 \Lc \de_\mu \de_\nu \pit$,
we have to check that $\tilde K_{\mu\nu} ^-$ is indeed the second derivative of a scalar field.
However this is automatic, since $\eta_{\mu\nu}$ itself is a second derivative,
$\eta_{\mu\nu} = \frac 1 2 \de_\mu \de_\nu  (x_\alpha x^\alpha)$.


\section{Classical stability}\label{stability}

We now  address in detail the issue of classical stability.
We will show that under very general conditions a solution which becomes trivial 
at spatial infinity is stable everywhere. 
Before proving this fact we want to specify what we mean by `stable'.
In particular, we are going to characterize the stability of a generic classical solution by studying
the tensor structure of the quadratic action of the fluctuations, eq.~(\ref{Z_stable}), around that
solution.
Notice that while for a stationary solution the concept of stability is perfectly defined,
for a solution which evolves in time such a definition is a bit tricky.
After all, if a solution depends on time the energy of the fluctuations around
that solution is not conserved.
For instance, one can expect to observe resonance phenomena associated to modes whose
proper frequency is of the order of the inverse time scale on which the
background solution evolves. A resonant growing fluctuation can be certainly seen as
an instability, but its presence is not simply encoded in the tensor structure of the quadratic action.
Our intention can therefore appear to be naive. 
However, the point is that our characterization is certainly sensible for ultraviolet modes,
well above all possible resonances, on time and length
scales much shorter than the typical time and length scales on which the background varies.
In such a regime it is perfectly acceptable to study the stability of the system as if $Z_{\mu\nu} (x)$
were constant in space and time.
We call this concept of stability `local', since it refers to the local structure of the quadratic
Lagrangian of the fluctuations, and in principle nothing prevents the system from being
stable in a given spacetime region (i.e.~stable against short wavelength fluctuations localized in that region),
and unstable in a different region.
In this section we prove that under proper although general conditions this cannot happen.

In a system described by a quadratic Lagrangian of the type of eq.~(\ref{Z_stable})  
two physically different kinds of instabilities must be considered.
The first is similar to a tachyonic instability, and, roughly speaking, is related to the relative signs of
the terms involving time and space derivatives. If one of the relative signs is wrong, there exists an exponentially 
growing mode.
More precisely, consider the equation of motion deriving from eq.~(\ref{Z_stable}),
\be
Z_{\mu\nu} \, \de^\mu \de^\nu \varphi = 0 \; ,
\ee
where, with the above discussion in mind, we neglected the space-time dependence of $Z_{\mu\nu}$.
For a mode with four-momentum $k^\mu = (\omega, \vec k)$, this gives the dispersion relation 
$Z_{\mu\nu} k^\mu k^\nu = 0$, which can be cast in the form
\be
(Z_{00} \, \omega + Z_{0i} \,k^i)^2 = (Z_{0i}Z_{0j} - Z_{00}Z_{ij}) k^i k^j \; .
\ee
In order for the system to be stable, the solution for $\omega$ must be real for any wave vector $\vec k$. 
The requirement is therefore that the 3$\times$3 matrix $Z_{0i}Z_{0j} - Z_{00}Z_{ij}$ is positive definite.
Although not manifestly, this condition is clearly Lorentz invariant, since it corresponds to the
requirement that the Lorentz invariant equation $Z_{\mu\nu} k^\mu k^\nu = 0$ has only real solutions 
(apart from an arbitrary overall phase multiplying $k^\mu$).

The second kind of instability one has to consider is the ghost-like instability, which is
related to the overall sign of the kinetic Lagrangian. 
Suppose that all relative signs are correct, so that all frequencies are real and there exists no 
exponentially growing mode. 
However, if the overall sign of the quadratic Lagrangian is wrong and our scalar field is coupled
to ordinary matter, then it is possible with zero total energy to amplify fluctuations in both sectors,
and this process is going to happen spontaneously already at the classical level.
At the quantum level the problem is probably even worse, although it is not yet clear how
to do sensible and consistent computations in a quantum theory with ghosts in the physical spectrum (see for instance \cite{me,cline}).
This kind of instability, unlike the previous, is there only in presence of interactions with
other sectors.
If the system is free of tachyonic instabilities, i.e.~if $Z_{0i}Z_{0j} - Z_{00}Z_{ij}$ is positive
definite, then the ghost-like instability is avoided if $Z_{00}>0$.

In the Hamiltonian formalism both types of instability, although physically different, are formally
on an equal footing.
Their absence is guaranteed whenever the Hamiltonian is positive definite.
For our action, eq.~(\ref{Z_stable}), the Hamiltonian density of the fluctuations reads
\be \label{H}
{\cal{H}}(\Pi,\varphi)=\frac{1}{4  Z_{00}}(\Pi + 2 Z_{0i} \, \nabla_i \varphi)^2 
- Z_{ij}  \, \nabla_i \varphi \nabla_j \varphi \; ,
\ee
where $\Pi$ is the conjugate momentum of $\varphi$, that is
$\Pi = 2(Z_{00} \, \dot \varphi - Z_{0i} \, \nabla_i \varphi)$.
${\cal{H}}$ is positive definite for positive $Z_{00}$ and negative definite $Z_{ij}$. 
The latter is a stronger requirement than the one we found above, 
since $Z_{0i}Z_{0j}$ is always a positive definite matrix.
However here there is a subtlety. What must be positive definite is not the Hamiltonian density, but
rather its integral in $d^3 x$, i.e.~the total Hamiltonian.
The mixed term $\sim \Pi \nabla_i \varphi$ that comes from expanding the square in eq.~(\ref{H})
is conserved on the equation of motion, once integrated in $d^3 x$, since its time derivative turns out to be a total gradient.
Therefore the integral
$H_0 = \int d^3 x {\cal{H}}_0$, with
\be
{\cal{H}}_0 (\Pi,\varphi)=\frac{1}{Z_{00}}\left[ \sfrac 14 \Pi^2 
+ (Z_{0i}Z_{0j} - Z_{00} Z_{ij})  \nabla_i \varphi \nabla_j \varphi \right] \; ,
\ee
is  conserved on the equations of motion of $H = \int d^3 x {\cal{H}}$. Although $H_0$ is not the Hamiltonian, a sufficient condition for stability 
(compact orbits) is that $H_0$ be positive definite. Notice that this condition is weaker than that set by positivity of $H$, but it coincides with the condition found from 
the field equations.
 In summary, in order for a solution to be stable one needs that $Z_{00}$ be positive
and that the matrix $Z_{0i}Z_{0j} - Z_{ij}$ be positive definite.


Consider now a generic distribution of matter sources for the $\pi$ field.
We stick to the case in which matter can be described by a fluid.
Its energy-momentum tensor is thus of the form $T_{\mu\nu}=(\rho+p) \, u_\mu u_\nu 
+ p \, \eta_{\mu\nu}$, where $\rho$ is the rest-frame energy density, $p$ is the rest-frame pressure, 
and $u^\mu$
is the fluid four-velocity.
The $\pi$ field is sensitive only to the trace $T = T^\mu {}_\mu = - (\rho - 3p)$,
which does not depend on $u^\mu$.
We consider the case in which the sources are localized, i.e.~their energy-momentum tensor decays
at spatial infinity.
In such a case far away from the source the trivial configuration $\tilde K_{\mu\nu}=0$, 
$Z_{\mu\nu}= -3 \eta_{\mu\nu}$ is a solution of the field equations.
Moreover, we start with the assumption that $p=0$ and $\rho \ge 0$, although we will relax it below.
We will show that under these hypotheses the system is stable.

Fix a generic $x^\mu$. $Z_{\mu\nu}$ is a ($x$-dependent) symmetric tensor, 
therefore we have some hope of diagonalizing it with a ($x$-dependent) Lorentz transformation.
This is not always possible, since the Minkowski metric $\eta_{\mu\nu}$ is not positive definite,
and the spectral theorem does not apply.
In the Appendix we show that $Z_{\mu\nu}$ can be certainly diagonalized with a Lorentz
transformation if the typical velocities of the matter sources are non-relativistic.
This is plausible from a physical point of view, since, roughly speaking, the failure of the
spectral theorem can be traced to a possible strong mixing between space and time that
can lead to a zero-norm eigenvector of $Z^\mu {}_\nu$.
For relativistic motions of the sources we are not guaranteed that the diagonalization is possible.
In order to proceed we have to assume it, so the result we are going to
find can be applied only if such a diagonalization is possible.
Let's therefore postulate that there exists
a Lorentz frame in which $Z^\mu {}_\nu = \mbox{diag}(z_0,z_1,z_2,z_3)$. 
In such a frame eq.~(\ref{Z_GC}) reads
\be \label{eigenvalues}
F(z) \equiv -\frac23 \left[(z_0^2 + \dots + z_3^2)-(z_0 z_1 + z_0 z_2 + \dots + z_2 z_3)\right] 
= 6 + \frac{\rho}{2 \Lc M_4} \ge 6 \; .
\ee
In order for the system to be locally stable in $x^\mu$ we want a positive $Z_{00}$ and a positive definite
$Z_{0i}Z_{0j} - Z_{ij}$. 
In the Lorentz frame we are using this simply reduces to the requirement that all $z_\mu$'s are negative.
In the space of the eigenvalues of $Z^\mu {}_\nu$ we can associate to this condition four critical 
hyperplanes, namely $z_\mu=0$, for $\mu=0, \dots, 3$. 
On these hyperplanes the system is marginally stable.
We want to show that these critical hyperplanes do not intersect the set of solutions of 
eq.~(\ref{eigenvalues}).
This result would permit us to conclude that the system is stable. 
In fact, at $\vec x \to \infty$ the system is locally stable, since the $\tilde K_{\mu\nu}=0$ 
configuration is stable. 
It could become unstable in moving to finite $\vec x$, but by continuity its trajectory in 
the eigenvalues space should cross one of the four critical hyperplanes. 
If the intersection between the critical hyperplanes and the set of solutions is null, this cannot happen.

The proof of the above fact is straightforward. Take a critical hyperplane, for instance $z_0 = 0$, 
and impose it as a constraint on the l.h.s.~of eq.~(\ref{eigenvalues}).
One gets
\begin{eqnarray}  
F(z) |_{z_0 = 0} & = & -\sfrac23 \left[(z_1^2 + z_2^2 + z_3^2)-(z_1 z_2 + z_1 z_3 + z_2 z_3)\right]\nonumber \\
               & = & -\sfrac13 \left[(z_1 - z_2)^2 + (z_1 - z_3)^2 + (z_2 - z_3)^2\right] \le 0
               \label{max} \; ,
\end{eqnarray}
while in order to solve eq.~(\ref{eigenvalues}) $F(z)$ should be larger or equal than 6.
In conclusion, the $z_\mu$'s that lie on a critical hyperplane cannot satisfy the field equation.

Of course, this argument does not prove that unstable solutions do not exist.
Actually, as we showed in the previous section, for every locally stable solution there 
exists a locally unstable conjugate solution. 
What this line of reasoning proves is that the sets of locally stable and locally unstable solutions are topologically 
disconnected in the eigenvalues space, and therefore a solution which is locally stable at one point 
(infinity, in our case) is stable everywhere. 
And vice-versa: a solution which is locally unstable at one point is unstable everywhere.
Notice that in the above proof the requirement that the sources are localized was needed only to ensure that at infinity
the trivial configuration $\tilde K_{\mu\nu}=0$ is a solution of the field equation, so that there exists
at least one point at which the system is locally stable.
In the case of non-localized sources, if a specific solution is known by other means to be locally stable at 
some point, then its stability throughout the Universe is guaranteed by the same arguments as above.

We have shown that, for sources with zero pressure and positive energy density decaying
at spatial infinity, the solution which is trivial at infinity is stable everywhere.
We can relax all these requirements, in order to find a condition under which the sets of locally stable
and locally unstable solutions in the eigenvalues space come into contact, so that a solution
could in principle be stable somewhere and unstable somewhere else.
Allowing for generic $\rho$ and $p$, eq.~(\ref{Z_GC}) now reads
\be
F(z) = 6 + \frac{(\rho - 3 p)}{2 \Lc M_4} \;.
\ee
From this and eq.~(\ref{max}) we see that a critical surface intersects the set of solutions if and only if
\be
(\rho - 3 p) \le -12 \, \Lc M_4 = -12 \, m^2 M_4 ^2 \;.
\ee
This is a necessary condition for the development of local instabilities, starting from a solution
locally stable at some point, for instance at infinity.

So far we have dealt with very general features of the classical solutions of the DGP model. We
now move to explicitly derive two interesting solutions and study their properties.


\section{de Sitter solution}\label{deSitter}

In refs.~\cite{deffayet, DDG} it was shown that the large-distance modification of gravity implied by the DGP model
can make a Friedmann-Robertson-Walker Universe  accelerate even in absence of a cosmological constant.
In particular, one finds that at late time, when all the energy density and pressure have redshifted away,
the Universe approaches an accelerating phase with constant Hubble parameter $H = m$. 
In other words, a solution of the DGP model is a four-dimensional de Sitter space with curvature radius $L=1/m$,
embedded in a five-dimensional Minkowski bulk.
The stability properties of this de Sitter solution were analyzed in ref.~\cite{LPR}, where it was shown that
around this configuration the $\pi$ field has a ghost-like kinetic term.

In this section we show that the four-dimensional effective action of the $\pi$ field, in absence of matter and
cosmological constant, can account for this de Sitter solution.
Of course, since de Sitter space is curved, while we are working in flat
space, we have to restrict to a small patch around $x=0$ and look at de Sitter space as a small deformation
of Minkowski space. This is sensible for $x \ll L$, where $L$ is the de Sitter curvature radius.

Consider eq.~(\ref{GC}) in vacuum, that is set $T=0$. 
We are looking for a maximally symmetric solution, therefore we consider the ansatz 
$\tilde K_{\mu\nu} = \sfrac 1 4 \tilde K \eta_{\mu\nu}$.
We obtain
\be
\tilde K = 4 \;, \qquad \tilde K_{\mu\nu} = \eta_{\mu\nu} \;, \qquad \pit (x) = -\sfrac 1 2 \Lc \, x_\mu x^\mu \; .
\ee
In the language of sect.~\ref{general}, this is the solution which is conjugate to the trivial 
configuration $\tilde K_{\mu\nu}=0$.
Therefore, we know that it must be unstable. 
The quadratic action of the small fluctuations $\varphi$ around this solution in fact is
\be
S_{ \varphi} = \int \! d^4 x \, \left[ - 3 (\de \varphi)^2 - 2 \left( \tilde K_{\mu \nu} - \eta_{\mu\nu} 
\tilde K \right) \de^\mu \varphi \de^\nu \varphi \right] = 
\int \! d^4 x \, 3 (\de \varphi)^2 \; ,
\ee
exactly reversed with respect to the `canonical' one.
This is precisely the result of ref.~\cite{LPR} about the kinetic term of the $\pi$ field around the 
de Sitter background.

To check that this solution corresponds locally to de Sitter geometry we must take into account the 
mixing between $\pi$ and the metric perturbation $h_{\mu\nu}$. 
This is straightforward to do by noticing that in terms of the Weyl-transformed metric 
$h'_{\mu\nu} = h_{\mu\nu} - \frac 1 {M_4} \pit \eta_{\mu\nu}$ the quadratic Lagrangian, eq.~(\ref{kinetic}), is
diagonal, i.e.~there is no mixing between $\pi$ and $h'_{\mu\nu}$.
The solution we are studying corresponds to $h'_{\mu\nu} = 0$, so that the
metric perturbation is
\be
h_{\mu\nu} = \frac 1 {M_4} \, \pit \eta_{\mu\nu} 
= -\sfrac 1 2 m^2 \, \eta_{\mu\nu} \, x_\alpha x^\alpha \; .
\ee
This describes locally a de Sitter space.
In fact, by performing a gauge transformation with parameter 
$\epsilon_\mu = \sfrac 1 4 m^2 \, x_\mu x_\alpha x^\alpha$, the metric becomes
\be
\tilde h_{\mu\nu} = h_{\mu\nu} + \de_\mu \epsilon_\nu + \de_\nu \epsilon_\mu = m^2 \, x_\mu x_\nu \; ,
\ee
and this is exactly the induced metric near $x=0$ on the hyperboloid
\be
\eta_{\mu\nu} \, x^\mu x^\nu + y^2 = \frac 1 {m^2} \, , 
\ee
which describes the embedding in five dimensional flat space of a four dimensional de Sitter space 
with curvature radius $L = 1/m = L_{\rm DGP}$.

We can check explicitly that in the regime of validity of the solution we have found
all higher order interactions, eq.~(\ref{general_int}), can be consistently neglected.
Our solution has $\pit \sim \Lc x^2$, $N'_\mu = 0$ and $h'_{\mu\nu} = 0$, so 
that eq.~(\ref{general_int}) reduces to
\be \label{higher_dS}
\Delta {\cal{L}}_{\rm bdy} \sim m M_4^2 \, \de (m \,x)^q \sim m^2 M_4^2 \, (m \,x)^{q-1} \; ,
\ee
to be compared with the cubic self-coupling of $\pi$, which on the solution is 
of order $\Lambda^6 \, x^2 = m^2 M_4^2 \, (m \,x)^2$.
We can trust our solution only for $x \ll L_{\rm DGP}=1/m$, since we are working in a nearly flat 
space. 
In this regime eq.~(\ref{higher_dS}) gives a negligible contribution 
to the action.


\section{Spherically symmetric solution}\label{spherical}

As pointed out in refs.~\cite{LS, DGZ}, one could in principle test the DGP model by solar system
experiments, at distances much smaller than the DGP scale, which instead is of the order of the Hubble horizon.
This is because the correction to the Newton potential of an isolated body 
becomes important at distances of the order of the so-called Vainshtein scale,
\be \label{vain}
R_V = \left( \frac{M}{m^2 M_4^2} \right)^{1/3} \sim \left( R_S L_{\rm DGP} ^2 \right)^{1/3} \ll L_{\rm DGP} \; ,
\ee
where $M$ is the mass of the body and $R_S$ its Schwarzschild radius.
At smaller distances the relative importance of this correction is small, of order $(r/R_V)^{3/2}$,
but nevertheless non negligible. 
An interesting fact is that, as shown in ref.~\cite{LS}, unlike its magnitude, the {\em sign} of this correction 
depends on the cosmological phase, i.e.~on the behaviour of the Universe at cosmological distances.

In our formalism all the relevant features of the DGP model are encoded in the self-interacting dynamics
of $\pi$.
We will derive an explicit solution for the field generated by a point-like source, and show that its 
properties agree with the above results.

Consider a static point-like source of mass $M$, located at the origin: $T = -M \, \delta ^3 (\vec x)$.
We look for a static spherically symmetric solution $\pit(r)$, where $r$ is the radial coordinates.
In such a case the field equations further simplifies; in particular eq.~(\ref{eq_pi}) becomes
\bea
\vec \nabla \cdot \left[6 \, \vec \nabla \pit - \frac 1 {\Lc} \vec \nabla (\vec \nabla \pit)^2 + \frac 2 {\Lc} 
\vec \nabla \pit \,\nabla^2 \pit \right] &  & \nonumber \\
= \vec \nabla \cdot \left[ 6 \, \vec E - \frac 1 {\Lc} \vec \nabla |\vec E|^2 + \frac 2 \Lc \vec E \, \vec \nabla \cdot 
\vec E   \right] &  & \nonumber \\
= \vec \nabla \cdot \left[
6 \, \vec E + \hat r \, \frac 4 \Lc \frac{E^2}{r} \right] & = & \frac {M}{2 M_4} \, \delta ^3 (\vec x) \; ,\label{radial}
\eea
where we defined $\vec E \equiv \vec \nabla \pit(r) \equiv \hat r E(r)$, and we used 
$\vec \nabla \cdot \vec E = \frac 1 {r^2} \frac d {dr} (r^2 \, E)$. 
After integration over a sphere centered at the origin, this reduces to an algebraic equation for $E(r)$,
\be
4 \pi r^2 \left( 6 E + \frac 4 \Lc \frac{E^2}{r}\right) = \frac {M}{2 M_4} \; ,
\ee
whose solutions are
\be
E_\pm (r) = \frac {\Lc}{4 r} \left[ \pm \sqrt{9 r^4+\sfrac 1 {2 \pi} \, R_V^3 \, r} - 3 r^2\right] \; ,
\ee
where $R_V$ is the Vainshtein scale, eq.~(\ref{vain}).
The solution for $\pit$ is obtained by integrating $E$ along $r$, $\pit_\pm (r)=\int E_\pm (r) dr$.

At small distances from the source the two solutions reduce to
\be
E_\pm (r \ll R_V) = \pm \frac {\Lc}{4 \sqrt{2 \pi}} \, \frac {R_V^{3/2}}{r^{1/2}} \; .
\ee
The acceleration induced on a test mass by $\pi$ is roughly $\frac 1 {M_4} E$. 
The relative correction to the Newton force is thus of order
\be
\frac{F_{\pi}}{F_{\rm Newton}} \sim \frac {E / M_4}{M / M_4 ^2 r^2} \sim \left( \frac r {R_V}  \right) ^{3/2} \; ,
\ee
as expected.
Notice that the magnitude of the correction does not depend on the solution chosen, but its sign does.
This fact matches to what we mentioned at the beginning of this section, namely that
the sign of the correction should depend on the behaviour of the Universe at cosmological distances.
Indeed, although the two solutions $E_\pm$ are similar to each other near the source, far away they behave very 
differently: $E_+$ decays as $1/r^2$, while $E_-$ blows up as $r^2$.
Since $E_+$ at spatial infinity reduces to the trivial configuration $\tilde K_{\mu\nu}=0$, 
one is tempted to conclude that $E_-$ should approach the de Sitter solution found in the 
previous section, since the de Sitter and the trivial solutions are conjugate to each other.
However, this is not the case, as one can easily check by noticing that $E_-$ does not reduce to a maximally
symmetric configuration at infinity. 
Indeed, $E_+$ and $E_-$ are not conjugate in the sense of sect.~\ref{general}, but rather
they are `spatially' conjugate, in the following sense.
$E_+$ is constant in time, therefore it produces a purely spatial $\tilde K_{\mu\nu} ^{+}$, that is
$\tilde K_{\mu 0} ^{+} = 0$.
By the same arguments of sect.~\ref{general} one easily concludes that another static solution of the
field equations is $\tilde K_{ij} =\frac 3 2 \delta_{ij} -\tilde K_{ij} ^{+}$.
$E_-$ corresponds precisely to this solution.

Since we are considering a localized source, with positive energy density, we know from
the results of sect.~\ref{stability} that the solution trivial at infinity, $E_+$, is stable
everywhere against small fluctuations.
We can check this fact explicitly.
For $\pit=\pit(r)$, eq.~(\ref{2nd_variation}) in spherical coordinates reads
\be
S_\varphi = \int \! d^4 x \sqrt g \, \left[ \left( 3 + \frac 2 \Lc \nabla^2 \pit \right)(\dot \varphi ^2 - g^{ij} 
\, \de_i \varphi \de_j \varphi)
+ \frac 2 \Lc \nabla_i \nabla_j \pit \, g^{ik} g^{jl} \, \de_k \varphi \de_l \varphi 
\right] \; ,
\ee
where $g_{ij} = {\rm diag} (1,r^2,r^2 \sin^2 \theta)$ is the metric tensor. 
The interesting non-zero components of the connection are
$\Gamma^r _{\theta\theta} = -r$ and $\Gamma^r _{\phi\phi} = -r \sin^2 \theta$, so that
\be
\nabla _r \nabla_r \pit = E' \; , \qquad \nabla _\theta \nabla_\theta \pit = r \, E\; , \qquad 
\nabla _\phi \nabla_\phi \pit = r\sin^2 \theta \, E \; ,
\ee 
while $\nabla^2 \pit = E' +\frac 2 r E$.
In the end we have
\bea
S_\varphi & = & \int \! d^4 x \sqrt g \left\{ 
\left[ 3 + \frac 2 \Lc \left( E' + \frac {2 \, E} r \right) \right] \dot \varphi^2  \right. \nonumber \\
 && \left. - \left[ 3 + \frac 4 \Lc \,\frac  E r \right] (\de_r \varphi)^2 -
\left[ 3 + \frac 2 \Lc \left( E' + \frac E r \right) \right] (\de_\Omega \varphi)^2
\right\} \; ,
\eea
where we defined $(\de_\Omega \varphi)^2$ as the angular part of $(\vec \nabla \varphi)^2$,
\be
(\de_\Omega \varphi)^2 \equiv (\vec \nabla \varphi)^2 - (\de_r \varphi)^2 =
\frac 1 {r^2} (\de_\theta \varphi)^2 + \frac 1 {r^2 \sin^2 \theta} (\de_\phi \varphi)^2 \;. 
\ee
By direct inspection  of the solution $E_+$ it is straightforward to see that the factors enclosed 
in square brackets are everywhere positive, thus proving the stability of the solution.

Similarly to the de Sitter case studied in the previous section, also on this solution higher order interactions 
of the type of eq.~(\ref{general_int}) give a negligible contribution to the action,
even well inside the Vainshtein region, where the extrinsic curvature is getting
larger and larger.
To see this, it is sufficient to plug into eq.~(\ref{general_int}) the behaviour of the solution for 
$R_S \ll r \ll R_V$, namely $\pit \sim \Lc R_V^{3/2}/r^{1/2}$, $h'_{\mu\nu} \sim R_S/r$ and $N'_\mu = 0$.
The result is
\be
\Delta{\cal{L}}_{\rm bdy} \sim \frac {m M_4^2}{R_S} \left(\frac{R_S} r \right)^{q/2 + s + 1} \; ,
\label{horizon}
\ee
which clearly shows that as long as $r \gg R_S$ the dominant contribution comes from the
cubic self-coupling of $\pi$, i.e.~from the term with $q=3$ and $s=0$.
Like in the de Sitter case, as long as the space is nearly flat the leading interaction term is
the cubic self-coupling of $\pi$, in agreement with the general argument given in sect.~\ref{effective}.

We want to conclude with a comment on the behaviour of the solution in the
presence of several sources. Of particular interest for the discussion of the following
section is the behaviour of the effective kinetic term coefficient $Z_{\mu\nu}$.
In the Vainshtein region $Z_{\mu\nu}$, the extrinsic curvature $K_{\mu\nu}$ and the Riemann tensor 
$R^\mu {}_{\nu\rho\sigma}$ scale as
\be\label{zriemann}
Z\,\sim\, K L_{\rm DGP}\,\sim\, \sqrt {R} \, L_{\rm DGP} \,\sim\, \sqrt {R_S} \, L_{\rm DGP}/r^{3/2} \; ,
\ee
where here on we neglect the indices. Notice, in particular, that the extrinsic and Riemann curvatures 
are associated to the same length scale: $K\sim \sqrt {R}$. For the asymptotic linear field
the relation was instead $K\sim R L_{\rm DGP}\ll {\sqrt{ R}}$. Now, the relation $Z\sim {\sqrt R} \, L_{\rm DGP}$
is valid also in more general situations. For instance we have checked it explicitly in the case
of a spherical matter distribution not localized at one point. It is also easy to derive it  for the interesting
case of two (several) pointlike sources, for instance the Sun and the Earth. As $M_\odot\gg M_\oplus$
we can neglect the Earth contribution in the asymptotic field. $M_\oplus$ becomes relevant only 
close enough to Earth. In this region we can approximately treat the Sun field as a background giving just rise
to a wave function $Z_\odot \sim \sqrt{M_\odot} \, L_{\rm DGP} /(M_4r_\oplus^{3/2})$ where $r_\oplus$ is the 
Earth-Sun distance.
By defining a canonical field $\pit_\odot=\sqrt{ Z_\odot} \pit$, the field equation close to Earth is just
eq.~(\ref{radial}) with renormalized values $\Lambda \to \Lambda_\odot\equiv \Lambda\sqrt{ Z_\odot}$,
$M_4\to M_4\sqrt{Z_\odot}$. Then applying the previous results we have that the renormalized Vainshtein
radius for the Earth field is ${R_\odot}_V^3=M_\oplus/(M_4 \Lambda^3 Z_\odot^2)=r_\oplus^3 M_\oplus/M_\odot$.
This is precisely the distance from Earth at which the Riemann curvature becomes dominated by the Earth
field. By applying the results of this section, at shorter distances the full $\pit$ wave function is just
$Z\sim Z_\odot ({R_\odot}_V/r)^{3/2}\sim \sqrt {R} \, L_{\rm DGP}$, where $R$ is now dominated by the Earth field. 
Notice in particular  that ${R_\odot}_V$ is larger than the Earth-Moon distance, so 
that the  effects of the $\pi$ field on the Moon
rotation \cite{LS,DGZ} can be studied by neglecting the presence of the Sun in first approximation. 
This is  precisely like in GR: the Sun's tidal force can be neglected when studying the Moon's orbit in lowest approximation.


\section{Quantum effects}\label{quantum}

Our discussion was until now classical. In order to characterize quantum versus classical 
effects it is useful to keep  $\hbar\not = 1$ and write the Lagrangian eq.~(\ref{action}) in terms of the non-canonical
 field
$\pi$ and the the 4 and 5D Newton's constants $G_4$ and $G_5$,
\be
{\cal{L}} =-\frac{G_4}{G_5^2}(\partial \pi)^2 +\frac{1}{G_5}(\partial \pi)^2 \partial^2\pi +\frac{G_4}{G_5}\pi T
\ee
(throughout this section we will neglect ${\cal O}(1)$ factors and be schematic with the contraction of
derivative and tensor indices).
 At the classical level physical quantities do not depend on the reparametrizations $\pi \to \sigma \pi$
and also ${\cal{L}}\to \eta {\cal{L}}$, with $\sigma, \eta$ constants. For instance, in the case of a localized source
$T=M\delta^3(x)$, the only invariant quantity is precisely the Vainshtein scale $R_V^3=M G_5^2/G_4$.
Indeed we can also define a classical expansion parameter, measuring the size of classical non 
linearities,
by taking the ratio of the second and first terms in the Lagrangian 
$\alpha_{cl}=G_5\partial ^2\pi/G_4\sim \tilde K$.
Notice that $\alpha_{cl}$ is invariant under the above reparametrizations. 
In the case of the spherical solution it is just a function of $r/R_V$. At the quantum level the
Lagrangian becomes a physical quantity, ${\cal{L}}\to \eta {\cal{L}}$ is no longer a `symmetry', so that
there is an additional physical length defined by  $L_Q^6=\hbar G_5^4/G_4^3$ or equivalently
an energy $\Lambda = \hbar /L_Q$. For $L_{\rm DGP}$ of the order
of the Hubble length we have $L_Q\sim 1000$ km. 
This analysis can be compared to 
the case of ordinary GR in 4D: the analogue of the Vainshtein scale is the Schwarzschild radius,
while the classical expansion parameter is simply the gravitational perturbation $h_{\mu\nu}=g_{\mu\nu}-
\eta_{\mu\nu}$. Similarly, at the quantum level the Planck length is the analogue of $L_Q$.

Like we have defined a classical expansion parameter $\alpha_{cl}$ we can define a quantum
expansion parameter $\alpha_q= (L_Q^2\partial\partial)$. In the case of a spherical
potential we have $\alpha_q=(L_Q/r)^2$. Then we expect that any physical quantity $A$ will be expandable
with respect to the value $A_0$ it takes in the linearized classical limit as a double series
\be
\frac{A-A_0}{A_0}=\sum_{n,m}a_{nm}\alpha_q^n \alpha_{cl}^m \; ,
\label{expansion}
\ee
with $n,m$ positive integers. The dependence on only integer powers of $\alpha_q$ follows from Lorentz
symmetry. As an example of a quantity of interest we could consider the action\footnote{Notice that although
the action is not a physical quantity at the classical level its relative variation $\delta S/S_0$ is
physical, and indeed defines $\alpha_{cl}$.}. As shown in ref.~\cite{LPR}, and as evident from eq.~(\ref{stable}),
the loops of the quantum fluctuation field $\varphi$  generate terms involving at least two
derivatives on the external background, i.e.~involving $\tilde K_{\mu\nu}=-\partial_\mu\partial_\nu \tilde \pi/\Lambda^3$
and its derivatives. 
In particular the tree level cubic interaction is not renormalized. 
\begin{figure}
\begin{center}
\includegraphics[width=9cm]{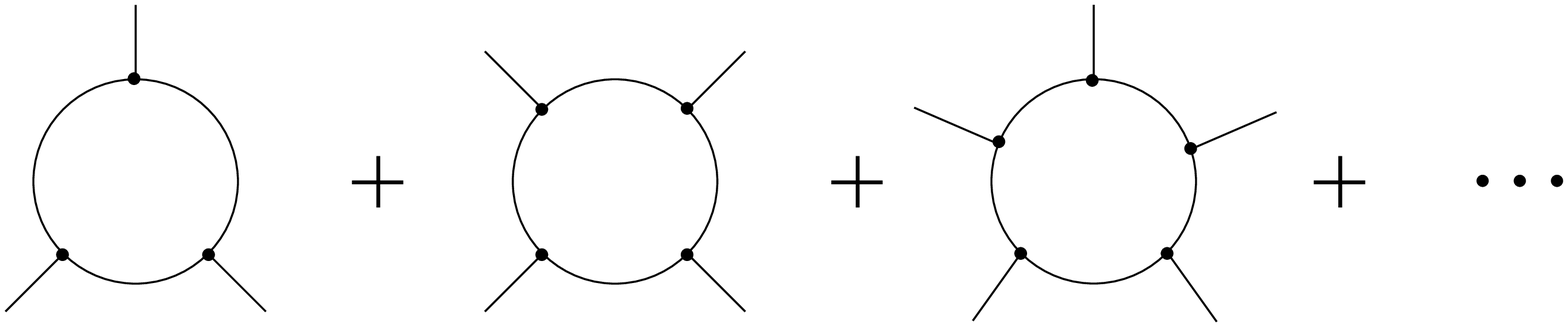}
\end{center}
\caption{The graphs contributing to the 1-loop effective action, eq.~(\ref{Gamma-1-loop}).\label{graphs}}
\end{figure}
For instance, cutting off the UV divergences at the scale $\Lambda$, the structure of the 1-loop correction is
\be \label{Gamma-1-loop}
\Gamma^{\rm 1-loop}=\sum_m\left [a_m\Lambda^4+b_m\partial^2\Lambda^2+\partial^4( c_m \ln \Lambda+I_m)\right]\left (\frac{\partial\partial \tilde \pi}{\Lambda^3}\right)^m  \; .
\label{1-loop}
\ee
The graphs contributing to $\Gamma^{\rm 1-loop}$ are depicted in fig.~\ref{graphs}; $m$ counts the number 
of external legs, and by $I_m$ we indicate the finite parts.  Notice 
that $\Gamma^{\rm 1-loop}/{\cal{L}}_{cl}= \Gamma^{\rm 1-loop}/(\partial\pi)^2$ has the structure of eq.~(\ref{expansion}). 
We should stress however that the power divergent terms are strictly speaking
not calculable, in the sense that they fully depend on the UV completion.
On the other hand the log divergent terms are also associated with infrared effects and are thus calculable. 
 They correspond in eq.~(\ref{expansion}) to powers of $n$ that are multiple of 3: $\alpha_q^3\sim 
\hbar\frac{G_5^4}{G_4^3} \partial^6= E^6/\Lambda^6$ is the genuine loop counting parameter.
For instance by working in dimensional regularization we would only get the minimal set of divergences that are
required for the consistent definition of the theory, and these are precisely the logarithmic ones.

Now the important remark is that when the classical parameter $\alpha_{cl}$ is larger than ${\cal{O}}(1)$ the expansion in 
eq.~(\ref{expansion})
 breaks down even if $\alpha_q\ll 1$. What basically happens is that large classical effects pump up the quantum
corrections, and the result becomes fully dependent on the series of counterterms.
Both power and log divergent terms at $\alpha_{cl}>1$ formally swamp the classical action
for a large enough number of external legs ($m$ in eq.~(\ref{expansion})).
This implies that the details of the UV completion of DGP in general dominate the physics for $\alpha_{cl}>1$ or 
at distances from a point source that are shorter than the Vainshtein radius $R_V$. Then 
 we should
not even be able to compute the gravitational potential at solar system distances, although they are still much bigger
than the genuine quantum length $L_Q$. 
A similar problem has been pointed out in the case of massive gravity \cite{AGS}.

There are however indications that the situation may  not be necessarily  as bad as it seems.
First, as we have already remarked, the problem arises when expanding in the number of external
legs, that is when expanding in the classical background field.
For any given amplitude with fixed number of legs the loop expansion, sum over $n$ in eq.~(\ref{expansion}),
is convergent as long as $r> L_Q$. 
Second the classical theory we are dealing with makes perfect sense
even below the Vainshtein scale as we have proven in the previous sections. For a well defined class of 
energy-momentum sources and of boundary condition at spatial infinity, no pathology or ghostlike
instability is encountered in the $\pi$ system. 
Indeed even if the interaction term involves four derivatives, the quadratic action around the background remarkably 
contains quadratic terms that have only up to two derivatives: this prevents the appearance of ghostlike modes. 
This should be compared to what happens when higher derivative corrections to a kinetic
term become important and ghost states appear: in those cases it is mandatory to assume that extra quantum effects
eliminate the problem, or better to declare the regime of validity of the theory to be limited. 
Based on these two remarks we want to consider the structure of the quantum effective action more closely.  
As we did previously we decompose $\pi$ into classical background and quantum fluctuation: $\pi=\pi_{cl}+\varphi$.  
The tree level Lagrangian for the fluctuation is
\be
{\cal L}_\varphi=Z_{\mu\nu}\partial^\mu\varphi\partial^\nu\varphi+\frac{1}{\Lambda^3}\partial^\mu\varphi\partial_\mu\varphi \Box \varphi
\label{flucttree} 
\ee
and $Z_{\mu\nu}$, see eq.~(\ref{def_Z}), contains all the dependence on the classical background $\pi_{cl}$. 
 In the regions where  classical non-linearities are large, the eigenvalues of the kinetic matrix become large, 
$Z \sim \tilde K \gg 1$. Now, as long as $Z$ does not lead to ghosts, a large $Z$ will enhance the gradient energy and suppress quantum fluctuations in $\varphi$.
This somewhat contradicts our previous conclusion that things are out of control when $\tilde K \gg 1$. 
 In order to simplify the discussion and the notation consider a toy model where the background matrix reduces to a scalar
 $Z_{\mu\nu}\equiv -\eta_{\mu\nu} Z$. As long as $Z$ varies slowly over space, it makes sense to define
a local strong interaction scale $\tilde \Lambda(x)=\Lambda \sqrt{Z(x)}$, which describes  the local scattering
of quantum excitations  $\varphi$. Of course this scale is obtained by
writing the action in terms of a locally canonical field $\hat \varphi=\varphi \cdot {\sqrt Z}$, 
and the definition makes sense
as long as the typical distance by which $Z$ 
changes is itself much bigger than $1/ \tilde \Lambda(x)$. 
It is easy to calculate the divergent part of the 1-loop effective action $\Gamma^{\rm 1-loop}$ by working
with the canonical field $\hat\varphi$. The dependence from $Z$ in $\Gamma^{\rm 1-loop}$ comes from two sources:
1) the Jacobian of the rescaling in the path integral,
2) a spacetime dependent  mass term $-m^2(x) \hat\varphi^2$ generated by the kinetic term after rescaling, 
\be
m^2(x)=\frac{1}{4}\left (\frac{\partial_\mu Z \partial^\mu Z}{Z^2}\right )-\frac{1}{2}\frac{\partial^2 Z}{Z} \; .
\label{mass}
\ee
The contribution from the mass term corresponds to the usual Coleman-Weinberg effective potential
 \be
 \Gamma_1^{\rm CW}= \frac{1}{16\pi^2}\left (\Lambda_{\rm UV}^2 m^2 -(m^2)^2 \ln \Lambda_{\rm UV}  +\hat\Gamma_1^{\rm CW}\right ) \; ,
\label{CW}
\ee
where the finite part $\hat\Gamma_1^{\rm CW}$ is expected to be a fairly complicated non-local expression that fully 
depends on the spacetime behaviour of $m^2$.
For instance, in  a central field where $Z\propto r^a$, like in the Vainshtein region, we have $m^2\sim 1/r^2$ and we 
expect $\hat\Gamma_1^{\rm CW}= -(m^2)^2\ln r$ plus terms of order $m^4$, but without log enhancement.
On the other hand the Jacobian factor yields formally
\be
\Gamma_1^{\rm J}=-\frac{1}{2}\delta^4(0) \ln Z \; .
\label{Jacobian}
\ee
 This term does not depend on the derivatives of $Z$, and should correspond to a quartic divergence.
 We can make this explicit by regulating the delta function as  $\delta^4(0) = \Lambda_{\rm UV}^4/(2\pi)^4$.
Notice that in principle the ultraviolet cutoff $\Lambda_{\rm UV}$ can be as high as the local scale $\tilde \Lambda(x)$.
To comment this results let us focus first on the logarithmic divergent term, as it is completely unambiguous. We see that when $Z \gg 1$
this term does not blow up, as one could have concluded by looking at the series expression eq.~(\ref{1-loop}). The effects of a large
$Z$ are resummed in the effective mass $m^2$, in a form that depends  only on the size of the relative derivatives $\partial^n Z/Z$. For instance
in the Vainshtein region the logarithmic part leads to a correction to the classical action which is of order
\be
\frac{1}{16\pi^2} \frac{1}{r^4} \ln (\Lambda_{\rm UV} r) \; .
\ee
This  should be compared with the classical action which is of order $\Lambda^4 (\Lambda R_V)^2 (R_V/r)^{5/2}=r^2 \tilde \Lambda^6(r)$: we find that 
the quantum correction is subdominant for $r>r_c=1/(\Lambda^4 R_V^3)=L_Q/(M_4 R_S)$. 
Indeed $r_c$ is precisely defined by the equality $r_c=\tilde \Lambda^{-1} (r_c)$ as one should have expected. Notice that for the interesting cases
of macroscopic sources $r_c$ is much shorter that $L_Q=1/\Lambda=1000$ km. We will have more to say on this below. We have more comments to make on this simple
1-loop computation. Eq.~(\ref{CW}), when expanded in powers of $\tilde K=Z-1$, corresponds to the calculation of the $c_m$ coefficients of eq.~(\ref{1-loop}).
This object describes the 1-loop RG evolution of the effective Lagrangian. On the other hand the initial conditions for this evolution at some scale,
for instance $\Lambda$, are determined by arbitrary local counterterms contained in the $I_n$. In general, as long  as locality and the symmetries are preserved,
there is little restriction on these terms from the effective Lagrangian view point. In particular they could become large just when $Z$ is large. For instance
a function of the form $Z^N (\partial Z/Z)^4$ would be a perfectly acceptable counterterm, with a perfectly fine analytic and local expansion around the point
$\tilde K= Z-1 = 0$. For $N \gg 1$ a term like this would dominate the action below the Vainshtein scale. What distinguishes a damaging term like this
from the mild effects that we have explicitly computed? The difference is very simple: this term depends on the absolute magnitude of $Z$, while the others don't.
After the rescaling, the dimensionful parameters in the tree level  $\hat\varphi$ Lagrangian are given by
 $\tilde \Lambda\equiv \Lambda {\sqrt Z}$ and by the derivatives
$\partial Z/Z\equiv 2 \partial \tilde\Lambda/\tilde\Lambda$,  $\partial^2 Z/Z\equiv\partial^2 \tilde \Lambda^2/\tilde\Lambda^2$, etc. The operator
in question cannot be written just in terms of these local scales since $Z^N= \tilde \Lambda^{2N}/\Lambda^{2N}$ and the original scale
$\Lambda$ does not influence local quantities in the tree approximation.  We can then characterize in a fairly simple way the class of 1-loop
counterterms with four derivatives (log divergences) for which perturbativity is preserved below the Vainshtein scale (in fact below $L_Q$):
it is given by local polynomials{\rm } of $\tilde \Lambda$, $\partial^2 \tilde \Lambda$, etc., weighted by the appropriate powers of $\tilde \Lambda$ to match dimensionality.
Nothing particularly interesting is learned by considering the structure of the power divergences. After all this is not surprising 
since they are totally scheme dependent. However the lesson we drew from the logarithmic ones is clear enough to allow us to formulate 
a simple consistent requirement, a  conjecture,  valid for all type of counterterms, and to all orders, such that the quantum corrections
remain under control way below the Vainshtein scale. The requirement is that the only dimensionful quantity determining the local counterterms
be the local scale $\tilde \Lambda(x)$,
\be
{\cal L}_{\rm CT}= \tilde \Lambda(x)^4 F\left (\frac{\partial }{\tilde \Lambda}\right) \; ,
\label{conjecture}
\ee
where $F$ schematically represents any infinite polynomial in $\partial \tilde \Lambda$, $\partial^2 \tilde\Lambda$, etc.
Notice that the above counterterm
Lagrangian should be added with $\tilde \Lambda$ written in terms of the physical field $\pi = \pi_{cl}+\varphi$, as this decomposition is arbitrary.
Then we need to check that eq.~(\ref{conjecture}) does not lead to new interactions for the quantum field $\varphi$ associated to scales lower than $\tilde \Lambda$.
If that were the case our ansatz would not be not self consistent since beyond 1-loop it would introduce a new scale in the effective action for the background field.
However it is easy to verify that this consistency check is satisfied. Notice that the minimal set of counterterms that are needed in DR with minimal subtraction
belongs to this class. However our definition is  more general. For instance a term with no derivative $\tilde \Lambda^4$ is consistent with 
eq.~(\ref{conjecture}) but is not generated in DR, since it corresponds to a quartic divergence.
The physical meaning of our ansatz is also clear. The scale $\tilde \Lambda$ truly represents the scale at which new degrees of freedom 
come in. The UV divergences in the low energy effective theory are cut off at this scale. For instance the vacuum energy is essentially
$\tilde \Lambda^4$ up to small derivative corrections associated to the space-time variation of $\tilde \Lambda$. From the point of view of the fundamental UV theory
the field $Z=1+\tilde K$ is a sort of dilaton that controls the mass gap $\tilde \Lambda$. At the points where $Z$ goes through zero 
the effective description breaks down since the gap vanishes and new massless states become relevant. This is another way to argue that
the points where $Z$ is negative, the ghost background, are outside the reach of our effective field theory and should be discarded.

We based all our discussion on the simplified case $Z_{\mu\nu}=-Z \eta_{\mu\nu}$, in which a rescaling of $\varphi$ 
allowed us to define the running scale $\tilde \Lambda (x)$. In the general case this cannot of course be done. The procedure then
is to work at each point with a short distance expansion of the $\varphi$ propagator,
\be
\langle \varphi(x+y)\varphi(x-y)\rangle =\frac{1}{Z_{\mu\nu}(x)y^\mu y^\nu}+d_1\frac{y^\rho y^\mu y^\nu \partial_\rho Z_{\mu\nu}(x)}{(yZy)^2}+\dots  \; ,
\ee
where the dimensionless coefficients (functions of $x$) $d_1,\dots$ are calculated in perturbation theory around a slowly varying $Z$ background. In spite of this more
complicated tensor structure the discussion is qualitatively the same: the $1/Z$ in the propagator arrange with the $1/\Lambda^3$ in the vertices
to yield a sort of running scale $\tilde \Lambda (x)$. However the dynamics at the scale $\tilde \Lambda$ will not be Lorentz invariant.
Anyway we stress that in the simple interesting case of a spherical source inside the Vainshtein region we have 
\be
Z_{\mu\nu} \simeq (R_V/r)^{3/2} A_{\mu\nu} \; ,
\ee
where $A_{\mu\nu}$ is a constant matrix. Then a simple rescaling of the quantum field is sufficient to study the effective action.

At this point one will wonder about naturalness. Are we advocating an incredibly tuned structure of
counterterms  not dictated by any symmetry but just by our desire to save the model? From the point of view of the effective theory expanded around
$\pi=0$ it looks like we are making a big tuning. This was the viewpoint of ref.~\cite{LPR} for DGP, and, earlier, of ref.~\cite{AGS}
for massive gravity. However, as we have proven in the first part of this paper, the validity of the classical theory
extends well beyond the point $\pi=0$. By validity of the classical theory here we include the fact that
$\tilde \Lambda$ never crosses $0$.
 All we have done in this section is to demand  uncalculable UV effects not to drastically perturb this picture throughout the patch
in field space where the classical theory works. The fact that this patch is sensibly bigger than the point $\pi=0$
leads to stronger constraints on the counterterms. But when looking at the region $\pi\not =0$ there is  more to say.
Around the points with $\partial\partial \pit \gg \Lambda^3$ our ansatz can indeed be motivated by a symmetry: scale invariance\footnote{
We thank Sergei Dubovsky for waking us up on this.}.
In the limit $\partial
\partial\pit\gg \Lambda^3$  we can neglect the quadratic kinetic term and work just with the cubic Lagrangian at tree level.
The cubic Lagrangian is obviously scale invariant by assigning $\pit$ dimension zero. We can also trivially absorb the 
powers of $\Lambda$ in the $\pit$ field: $\pit=\Lambda \tilde \pi$. The background $\tilde \pi_{cl}$  spontaneously breaks scale invariance.
In particular it generates the mass scale $\tilde \Lambda^2\sim \partial\partial \tilde\pi_{cl}$.  
Then it is obvious that our ansatz preserves scale invariance. The original quadratic kinetic term now reads 
$\Lambda^2 (\partial \tilde \pi)^2$ representing a soft explicit breaking of scale invariance. When treating this term as a small perturbation
it is then natural to
 generalize our eq.~(\ref{conjecture}) to include all the terms proportional to positive powers of $\Lambda^2/{\tilde\Lambda}^2$.
Now, one should be aware that scale invariance, unlike other global symmetries, cannot be 
readily used  to naturally enforce parameter choices. Basically this is because conformal invariance in the interesting cases is
 never exact in the far UV.
For instance quantum gravity breaks it. This limits the use of conformal symmetry as a substitute of supersymmetry in attacking
the gauge hierarchy problem (see for instance a discussion in ref.~\cite{zaffa}). However one can imagine a situation
where the scale $M$ at which a models flows to a conformal point is much smaller than the physical and conformal breaking cut-off $\Lambda_{\rm UV}$.
Then even though at the fixed point there may exist relevant deformations their coefficient will be naturally small (in units of $M$)
if the corresponding operator is irrelevant at the UV scale. An example could be a 4-fermion term suppressed by $1/\Lambda_{\rm UV}^2$ 
in the microscopic
 theory, which becomes relevant at the fixed point thanks to a big anomalous dimensions. 
Such an  example suggests that a theory may remain naturally scale invariant in a wide range of scales
below the (conformal breaking) cut-off.
 In our case conformal symmetry is surely broken at the scale $M_5$, but we do not need to go that high in energy. This keeps open
the possibility that our ansatz may indeed be technically natural. It would be interesting to investigate these issues in more
detail, perhaps trying to come up with an explicit realization of a conformally invariant completion.


Finally we want to discuss the physical implications of our results. The running scale close to a spherical source can also be written as
\be
\tilde \Lambda (r)= M_5^{\frac{1}{2}}R_S^{\frac{1}{4}} r^{-\frac{3}{4}} \; ,
\label{running}
\ee
from which the critical length $r_c$ at which the classical background $\pi_{cl}$ stops making sense is just
$r_c=1/R_S M_5^2$. We have $r_c< R_S$ for $R_S>1/M_5\sim 1$ fm, so that the quantum effects in the field $\pi$ have
no implication on any relevant astrophysical black hole. Of course we still have to study the classical effects of
the field $\pi$ on 4D black holes.

In the presence of several macroscopic sources, like in the solar system, eq.~(\ref{running}) is dominated by the one
generating the biggest gravitational tidal effect, i.e.~the one dominating the Riemann tensor $\sim R_S/r^3$. 
At the surface of the Earth it is the Earth field itself that dominates $\tilde \Lambda$ and we find
\be
\tilde \Lambda_\oplus ^{-1} \sim 1 \: {\rm cm}  \; .
\ee
In principle this is the length scale below which we cannot compute the gravitational potential at the surface of the Earth!
Notice that this scale exceeds by almost 2 orders of magnitude the present experimental bound on modifications
of gravity \cite{eotwash,kapitulnik,LP}. 
Notice also that we cannot play  with $M_5$ in order to increase $\tilde \Lambda_\oplus$ by $10^2$.
This is because $ \tilde \Lambda_\oplus \propto \sqrt M_5$ while $L_{\rm DGP}\propto 1/M_5^3$ so that we would 
lower at the same time
$L_{\rm DGP}$ down to the unacceptably low value of $\sim 1$ kpc.
However the 1 cm scale we are discussing is the one at which $\pi$, not the usual graviton, goes into a quantum fog.
The graviton at this scale is still very weakly coupled. Moreover the field $\pi$ at distances just a bit bigger that 
1 cm is a very small source of macroscopic gravity, its contribution relative to the ordinary one being of order
\be
\left ( \frac{R_\oplus}{R_V}\right )^{\frac{3}{2}}\equiv \frac{\Lambda^2}{\tilde \Lambda_\oplus^2}\sim 10^{-16} \; .
\ee
Then it is reasonable to expect that whatever UV physics will take over below 1 cm it will not raise the effects
of this sector on the gravitational potential by 16 orders of magnitude right away. Nonetheless this result
may be viewed as yet another motivation to study the gravitational forces below a mm.


\section{Outlook}

We conclude by briefly mentioning some directions in which our analysis could be extended.
First of all, there is the issue of stability of time-dependent classical solutions. 
As we have already stressed, our proof of classical stability is rigorously valid only
for matter sources with non relativistic velocities. It would be interesting to understand to
what extent the stability is guaranteed in the presence of relativistic sources.
Perhaps one can encounter astrophysical or cosmological situations in
which the kinetic Lagrangian of the fluctuation around the background
passes through zero in some spacetime region. Such a region would be
characterized by a local scale $\tilde \Lambda = 0$. This clearly
indicates a breakdown of the effective theory, and in principle one
expects unsuppressed quantum effects to take over, with interesting
phenomenological signatures. Of course these are, by definition, out of
the reach of the effective theory, and we can say nothing about them
without a UV completion of the model.

On the other hand it would also be interesting to study whether the
approximate conformal symmetry of our ansatz can tell us more
about the UV completion. Notice in passing that by neglecting the original
quadratic kinetic
term, the $\pi$-sector becomes a  critical\footnote{
The kinetic term is not negative but zero.} ghost
condensate model. It is the presence of
a classical energy-momentum source,
 and not the $\pi$ self-interaction,  that
causes the $\pi$ to condense, giving thus
rise to a well defined effective field theory.

Another point which may require extra study concerns 4D black holes. Are they  different from those of ordinary GR?
The fact that the field $\pi$, when approaching the Schwarzschild radius, gives a more and more negligible
correction to the Newton force suggests that black holes should be unaffected by its presence.
However the very expansion  we use breaks down close to the Schwarzschild radius, as shown by eqs.~(\ref{general_int}, \ref{horizon}).
Therefore a (numerical) study  of the exact solution may be required. Of course the problem we are encountering here
is just that close to the horizon we cannot consider the brane as approximately flat. A better viewpoint
is gained by considering the full Einstein equations at the boundary in eq.~(\ref{boundaryem}). In the region $R_S\ll r\ll R_V$  the extrinsic
curvature term, which controls deviations from GR, represents a small perturbation. This can be phrased as
\be
R(g)\sim \sqrt {R_{\mu\nu}R^{\mu\nu}}\sim  \frac{K}{L_{\rm DGP}}=\frac{1}{L_{\rm DGP}}\sqrt{\frac{R_S}{r^3}}\ll \sqrt{R_{\mu\nu\rho\sigma}
R^{\mu\nu\rho\sigma}}\sim \frac{R_S}{r^3} \; ,
\ee
corresponding to the fact that the geometry is approximately Schwarzschild. The only way this state of things can drastically change is
if the extrinsic curvature develops a singularity while approaching the Schwarzschild region. Without an explicit computation
we cannot exclude this possibility, but we find it  very unlikely. In analogy with GR,  the horizon should not 
be a special locus from the point of view of the curvature.

Finally, it would be worth analyzing in more detail the connection between DGP and massive gravity,
and investigating the possibility that our results, or some of them, can be exported to the latter.
This is far from trivial, since massive gravity apparently doesn't share with DGP two remarkable features that turned
out to be crucial in our analysis:
the fact that the equation of motion for the strongly interacting field depends {\em algebraically}
on the coefficients of the quadratic action around the background, and the absence in that action of terms with 
more than two derivatives.

\section*{Acknowledgements}

We would like to thank Nima Arkani-Hamed, Sergei Dubovsky, Gregory Gabadadze, Massimo Giovannini, Pietro Grassi, 
Thomas Gregoire, Arthur Lue, Markus Luty, Massimo Porrati, Valery Rubakov and Arkady Vainshtein for useful discussions.
A.~N.~is supported by EU under the RTN contract HPRN-CT-2000-00152.

\section*{Note added}

In a recent paper ref.~\cite{dvali} it was also argued that
the UV completion of DGP should not destroy predictivity.
Such a claim  is in the same spirit of our paper, although ref.~\cite{dvali}
is mostly  based on an interesting $\sigma$-model version of DGP, where there are just scalar fields.
In that toy model there is no analogue of our $\pi$ field, with all
its associated peculiarities, so that it is not immediately clear to us
that there is a direct connection to our work. It would be interesting
to further investigate this possibility.


\appendix
\section{Diagonalization of $Z_{\mu\nu}$ in the non-relativistic limit}

In this Appendix we want to prove that whenever the matter sources have non-relativistic velocities
the tensor $Z_{\mu\nu}$ is diagonalizable with a Lorentz transformation.

First, from the very definition of $Z_{\mu\nu}$, eq.~(\ref{def_Z}), we see that  
if $\tilde K_{\mu\nu}$ is diagonalizable, so is $Z_{\mu\nu}$. 
Given $\tilde K_{\mu\nu}$ in a generic frame, by means of a spatial rotation we
can always align the spatial vector $\tilde K_{0i}$ to the $x$ axis and cast $\tilde K_{\mu\nu}$ 
in the form
\be
\tilde K_{\mu\nu} = \left( \begin{array}{cccc}
	k_0 & k_{01} & 0 & 0\\
	k_{01} & k_1 & \star & \star\\
	0 &  \star& \star & \star  \\
	0 & \star & \star & \star
\end{array}\right) \; .
\ee
Now, the point is that if by means of a boost along $x$ we succeed in diagonalizing the upper-left
2$\times$2 block we are done, since then by a spatial rotation we can diagonalize the full matrix.
The problem is thus reduced to the diagonalization of a 2$\times$2 symmetric tensor with a boost,
i.e.~to find two real eigenvectors of the matrix
\be
\tilde K^\alpha {}_\beta = \left( \begin{array}{rc}
	-k_0 & -k_{01} \\
	k_{01} & k_1 \\
	\end{array}\right) \; ,
\ee
orthonormal with respect to the 2D Minkowski metric $\eta_{\alpha\beta}$ 
(now the indices $\alpha$ and $\beta$ run over 0 and 1).
The eigenvalues of $\tilde K^\alpha {}_\beta$ are
\be \label{eigen_lambda}
\lambda_\pm = \sfrac12 (k_1-k_0) \pm \sfrac12 \sqrt{(k_1+k_0)^2 - 4 k_{01}^2 } \; ,
\ee
which are real and distinct as long as the off-diagonal element is small, $|k_{01}|< \frac12 |k_1+k_0|$.
In such a case there exist two real eigenvectors, orthonormal with respect to $\eta_{\alpha\beta}$.
When instead $|k_{01}| \to \frac12 |k_1+k_0|$ the two eigenvalues become equal, there is only one eigenvector
and it has zero norm. Beyond this point, when $|k_{01}| > \frac12 |k_1+k_0|$, both the eigenvalues and 
the eigenvectors are complex.

Now we want to make contact with the physics.
The tensor $\tilde K_{\mu\nu}$ is related to the second derivatives of $\pi$, 
$\tilde K_{\mu\nu} = -\frac 1 \Lc \de_\mu \de_\nu \pi$, so that
\be
k_0 = -\sfrac 1 \Lc \ddot \pi \; , \quad 
k_{01} = -\sfrac 1 \Lc \de_x \dot \pi \; , \quad 
k_1 = -\sfrac 1 \Lc \de_x^2 \pi \; .
\ee
If the matter sources are stationary we can certainly choose $\pi$ to be independent on time.
In such a case $k_0 = k_{01} = 0$.
Now we can look at the case in which the sources are evolving with non-relativistic velocities as
a perturbation of the stationary case.
In particular, if the typical velocity of the sources is $v$, then we expect
that the time derivatives of $\pi$ will be suppressed by powers of $v$ with respect to the
spatial gradients, namely
\be
k_0 \sim v^2 \, k_1 \;, \qquad k_{01} \sim v \, k_1  \; .
\ee
In the non-relativistic case, $v \ll 1$, the two eigenvalues in eq.~(\ref{eigen_lambda})
are certainly real and distinct,
\be
\lambda_+ = k_1 - \frac{k_{01}^2}{k_1} + {\cal{O}}(v^3 \, k_1) \; , \qquad
\lambda_- = -k_0 + \frac{k_{01}^2}{k_1} + {\cal{O}}(v^3 \, k_1) \; .
\ee
This shows that for non-relativistic sources the 2$\times$2 matrix $\tilde K^\alpha {}_\beta$
is diagonalizable with a Lorentz boost, so that the full 4$\times$4 $\tilde K_{\mu\nu}$
is diagonalizable, and so is $Z_{\mu\nu}$.



\begin{thebibliography}{99}

\bibitem{PF}
M.~Fierz and W.~Pauli,
``On Relativistic Wave Equations For Particles Of Arbitrary Spin In An Electromagnetic Field,''
Proc.\ Roy.\ Soc.\ Lond.\ A {\bf 173}, 211 (1939).

\bibitem{van} 
P.~Van Nieuwenhuizen,
``On Ghost-Free Tensor Lagrangians And Linearized Gravitation,''
Nucl.\ Phys.\ B {\bf 60}, 478 (1973).

\bibitem{GRS}
R.~Gregory, V.~A.~Rubakov and S.~M.~Sibiryakov,
``Opening up extra dimensions at ultra-large scales,''
Phys.\ Rev.\ Lett.\  {\bf 84}, 5928 (2000)
[arXiv:hep-th/0002072].

\bibitem{DGP}
G.~R.~Dvali, G.~Gabadadze and M.~Porrati,
``4D gravity on a brane in 5D Minkowski space,''
Phys.\ Lett.\ B {\bf 485}, 208 (2000)
[arXiv:hep-th/0005016].

\bibitem{ACLM}
N.~Arkani-Hamed, H.~C.~Cheng, M.~A.~Luty and S.~Mukohyama,
``Ghost condensation and a consistent infrared modification of gravity,''
arXiv:hep-th/0312099.

\bibitem{Dvali:2002pe}
G.~Dvali, G.~Gabadadze and M.~Shifman,
``Diluting cosmological constant in infinite volume extra dimensions,''
Phys.\ Rev.\ D {\bf 67}, 044020 (2003)
[arXiv:hep-th/0202174].

\bibitem{deffayet}
C.~Deffayet,
``Cosmology on a brane in Minkowski bulk,''
Phys.\ Lett.\ B {\bf 502}, 199 (2001)
[arXiv:hep-th/0010186].

\bibitem{DDG}
C.~Deffayet, G.~R.~Dvali and G.~Gabadadze,
``Accelerated universe from gravity leaking to extra dimensions,''
Phys.\ Rev.\ D {\bf 65}, 044023 (2002)
[arXiv:astro-ph/0105068].

\bibitem{LS}
A.~Lue and G.~Starkman,
``Gravitational leakage into extra dimensions: Probing dark energy using local gravity,''
Phys.\ Rev.\ D {\bf 67}, 064002 (2003)
[arXiv:astro-ph/0212083].

\bibitem{DGZ}
G.~Dvali, A.~Gruzinov and M.~Zaldarriaga,
``The accelerated universe and the Moon,''
Phys.\ Rev.\ D {\bf 68}, 024012 (2003)
[arXiv:hep-ph/0212069].

\bibitem{LPR}
M.~A.~Luty, M.~Porrati and R.~Rattazzi,
``Strong interactions and stability in the DGP model,''
JHEP {\bf 0309}, 029 (2003)
[arXiv:hep-th/0303116].

\bibitem{rub}
V.~A.~Rubakov,
``Strong coupling in brane-induced gravity in five dimensions,''
arXiv:hep-th/0303125.

\bibitem{v}
A.~I.~Vainshtein,
``To The Problem Of Nonvanishing Gravitation Mass,''
Phys.\ Lett.\ B {\bf 39}, 393 (1972).

\bibitem{ddgv}
C.~Deffayet, G.~R.~Dvali, G.~Gabadadze and A.~I.~Vainshtein,
``Nonperturbative continuity in graviton mass versus perturbative discontinuity,''
Phys.\ Rev.\ D {\bf 65}, 044026 (2002)
[arXiv:hep-th/0106001].

\bibitem{AGS}
N.~Arkani-Hamed, H.~Georgi and M.~D.~Schwartz,
``Effective field theory for massive gravitons and gravity in theory space,''
Annals Phys.\  {\bf 305}, 96 (2003)
[arXiv:hep-th/0210184].

\bibitem{vdvz}
H.~van Dam and M.~J.~Veltman,
``Massive And Massless Yang-Mills And Gravitational Fields,''
Nucl.\ Phys.\ B {\bf 22}, 397 (1970); 
V.~I.~Zakharov, JETP \ Lett. {\bf 12},
312 (1970).

\bibitem{g}
A.~Gruzinov,
``On the graviton mass,''
arXiv:astro-ph/0112246.

\bibitem{p3}
M.~Porrati,
``Fully covariant van Dam-Veltman-Zakharov discontinuity, and absence  thereof,''
Phys.\ Lett.\ B {\bf 534}, 209 (2002)
[arXiv:hep-th/0203014].

\bibitem{Tanaka}
T.~Tanaka,
``Weak gravity in DGP braneworld model,''
Phys.\ Rev.\ D {\bf 69}, 024001 (2004)
[arXiv:gr-qc/0305031].

\bibitem{GH}
G.~W.~Gibbons and S.~W.~Hawking,
``Action Integrals And Partition Functions In Quantum Gravity,''
Phys.\ Rev.\ D {\bf 15}, 2752 (1977); 
S.~W.~Hawking and G.~T.~Horowitz,
``The Gravitational Hamiltonian, action, entropy and surface terms,''
Class.\ Quant.\ Grav.\  {\bf 13}, 1487 (1996)
[arXiv:gr-qc/9501014]. 

\bibitem{me} 
R.~Rattazzi,  
``A new dimension at ultra large scales and its price'', 
talk at SUSY2K, http://wwwth.cern.ch/susy2k/susy2kfinalprog.html, unpublished.

\bibitem{cline}
J.~M.~Cline, S.~y.~Jeon and G.~D.~Moore,
``The phantom menaced: Constraints on low-energy effective ghosts,''
arXiv:hep-ph/0311312.

\bibitem{zaffa}
R.~Rattazzi and A.~Zaffaroni,
``Comments on the holographic picture of the Randall-Sundrum model,''
JHEP {\bf 0104}, 021 (2001)
[arXiv:hep-th/0012248].

\bibitem{eotwash}
C.~D.~Hoyle, U.~Schmidt, B.~R.~Heckel, E.~G.~Adelberger, J.~H.~Gundlach, D.~J.~Kapner and H.~E.~Swanson,
``Sub-millimeter tests of the gravitational inverse-square law: A search  for `large' extra dimensions,''
Phys.\ Rev.\ Lett.\  {\bf 86}, 1418 (2001)
[arXiv:hep-ph/0011014];
E.~G.~Adelberger  [EOT-WASH Group Collaboration],
``Sub-millimeter tests of the gravitational inverse square law,''
arXiv:hep-ex/0202008.

\bibitem{kapitulnik}
J.~Chiaverini, S.~J.~Smullin, A.~A.~Geraci, D.~M.~Weld and A.~Kapitulnik,
``New experimental constraints on non-Newtonian forces below 100 $\mu$m,''
Phys.\ Rev.\ Lett.\  {\bf 90}, 151101 (2003)
[arXiv:hep-ph/0209325].

\bibitem{LP}
J.~C.~Long, H.~W.~Chan and J.~C.~Price,
``Experimental status of gravitational-strength forces in the  sub-centimeter regime,''
Nucl.\ Phys.\ B {\bf 539}, 23 (1999)
[arXiv:hep-ph/9805217];
J.~C.~Long and J.~C.~Price,
``Current short-range tests of the gravitational inverse square law,''
Comptes Rendus Physique {\bf 4}, 337 (2003)
[arXiv:hep-ph/0303057].

\bibitem{dvali}
G.~Dvali,
``Infrared modification of gravity,''
arXiv:hep-th/0402130.



\end{thebibliography}
\end{document}